\documentclass[]{spie}  
\usepackage[]{graphicx}

\def\sci#1{{\; \times \; 10^{#1}}}
\def\depang{64\,degrees}
\def\fov{$4.7$\,arcminute}

\newenvironment{my_enumerate}{
\begin{enumerate}
  \setlength{\itemsep}{1pt}
  \setlength{\parskip}{0pt}
  \setlength{\parsep}{0pt}}{\end{enumerate}
}

\title{Detailed Design of a Deployable Tertiary Mirror for the Keck I Telescope} 

\author{
J. Xavier Prochaska\supit{a},
Chris Ratliff\supit{a},
Jerry Cabak\supit{a}, 
Alex Tripsas\supit{a},
Sean Adkins\supit{b},
Michael Bolte\supit{a},
David Cowley\supit{a},
Mike Dahler\supit{b},  
Will Deich\supit{a},  
Hilton Lewis\supit{b}, 
Jerry Nelson\supit{a}, 
Sam Park\supit{b},     
Michael Peck\supit{a}, 
Drew Phillips\supit{a},
Mike Pollard\supit{b}, 
Bill Randolph\supit{b},
Dale Sanford\supit{a}, 
Jim Ward\supit{a},     
Truman Wold\supit{b} 
\skiplinehalf
\supit{a}University of California Observatories, 1156 High St., Santa Cruz, CA 95064 USA; \\
\supit{b}W.M. Keck Observatory, 65-1120 Mamalahoa Hwy, Kamuela, HI 96743, USA
}

\begin{document} 
  \maketitle 

\begin{abstract}
Motivated by the ever increasing pursuit of science with the
transient sky (dubbed Time Domain Astronomy or TDA), we are
fabricating and will commission a new deployable tertiary mirror
for the Keck I telescope (K1DM3)
at the W.M. Keck Observatory.  This paper presents the detailed
design of K1DM3 with emphasis on the opto-mechanics.  
This project has presented several design challenges.
Foremost are the competing requirements to avoid vignetting
the light path when retracted against a sufficiently rigid
system for high-precision and repeatable pointing.
The design utilizes an actuated swing arm to retract the
mirror or deploy it into a kinematic coupling.
The K1DM3 project has also required the design and development
of custom connections to provide power, communications,
and compressed air to the system.  This NSF-MRI funded
project is planned to be commissioned in Spring 2017.
\end{abstract}



\section{Introduction}
\label{sec:intro}  

A major thrust of astronomy in the 21st century is to study, observationally and with theoretical inquiry, time-variable phenomena in the night sky.  This area is broadly referred to as time domain astronomy (TDA) and its high scientific priority was established by the Astro2010 report (National Research Council, 2010).  Their highest recommendation for large telescope ground-based observing, for example, was to build the Large Synoptic Survey Telescope (LSST).  In advance of that ambitious project, several projects are using wide-field cameras to image large areas of the sky at high cadence.  This includes the partially NSF-sponsored Palomar Transient Factory (PTF)
and its NSF/MSIP funded follow-on (ZTF),
 and the Pan-STARRS surveys which repeatedly image the full northern sky, finding hundreds of new transient phenomena on every clear night. These surveys are discovering thousands of supernovae, immense samples of asteroids and near-Earth objects, variable stars of diverse nature, flaring phenomena, and other exotic sources. These advances in TDA observing at optical wavelengths follow decades of TDA science performed at higher energies from space.  Indeed, the first astronomical sources detected with $\gamma$-rays were themselves transient phenomena: the so-called $\gamma$-ray bursts (GRBs). Satellites like NASA's Swift and Fermi monitor $\approx \pi$ steradians of the sky, scanning for transient and variable high-energy events.

The focus of most previous and on-going TDA projects has been wide-field imaging of the sky in search of rare and new classes of events.  To fully explore and exploit the astrophysics of newly discovered sources, however, one must establish the redshift and/or the type of object responsible. Optical and infrared wavelengths remain the most powerful and efficient passbands to perform the required spectroscopy.  This is the primary role of large, ground-based observatories in TDA science.  Recognizing their value, several 8\,m-class observatories have established very effective observing strategies (generally at great expense) to perform such science.  Both the Gemini and European Southern Observatories (ESO) designed their largest telescopes with systems that could rapidly feed any of the available foci.  Furthermore, they designed queue operations to enable rapid responses to targets-of-opportunity (ToOs) and programs that repeatedly observe a source for short intervals at high cadence.  Successes of this model include time-resolved spectroscopy of varying absorption lines from a GRB afterglow on minute time-scales and high-cadence monitoring of the Galactic center to recover high fidelity orbital parameters.

The W.M. Keck Observatory (WMKO)
boasts twin 10\,m telescopes, currently the largest aperture, fully-operational optical/IR telescopes.  Over the course of the past ~20 years, we have successfully instrumented each telescope with high-throughput imagers and spectrometers spanning wavelengths from the atmospheric cutoff to several microns. The Keck I (K1) telescope hosts the Nasmyth-mounted HIRES spectrograph, one of the primary tools for obtaining high-resolution visible wavelength spectra from TDA observations. K1 also hosts the Nasmyth-mounted Keck I adaptive optics (AO) system with a high-performance
laser guide star (LGS) system, and now hosts the near-IR integral field spectrograph OSIRIS, a key tool for synoptic observations of the Galactic center. Two K1 instruments are used at Cassegrain: the LRIS multi-object visible wavelength spectrograph and the near-IR multi-object spectrograph and imager MOSFIRE. This is a unique instrument suite, especially within the U.S. community: the HIRES spectrometer is the only echelle spectrometer on a large aperture telescope in the northern hemisphere; LRIS provides extremely sensitive spectroscopy especially at blue ($< 4000$\AA) and red ($> 8000$\AA) wavelengths, and MOSFIRE represents a unique capability for multi-slit, near-IR spectroscopy in the northern hemisphere.

By its nature, TDA science demands a more nimble and flexible approach to observations than the traditionally, classically-scheduled observing which has been the standard at WMKO. Most TDA programs require observations made with a specific instrument at specific times, while classical scheduling on a telescope with multiple instrument configurations may mean that the desired instrument will not be available for the TDA program.
In the current configuration of the Keck telescopes a removable module, called the tertiary module, which contains the telescope tertiary mirror (M3), is used to support observations with Nasmyth and bent Cassegrain mounted instruments. 
The desired instrument along the elevation axis ring is selected by rotating the tertiary mirror around the telescope optical axis. To install and use a Cassegrain mounted instrument, the tertiary module must be removed from the telescope which is a process typically requires
daycrew.

The K1 deployable tertiary (K1DM3) will directly address these shortcomings, enabling WMKO to fully and vigorously participate in TDA science with its unique set of K1 instruments.  As the driving paradigm in observational astronomy shifts from a passive, static sky to one that displays dramatic changes on a nightly basis, it is critical to enhance our technical capabilities in this arena.
The K1DM3 will increase the flexibility for ToO and Cadence observations with the Nasmyth, bent Cassegrain, and whichever Cassegrain instrument is installed in the telescope, without requiring any configuration change other than rotating the tertiary mirror to the appropriate focal station or retracting the mirror from the telescope beam. The K1DM3 will also reduce the time required for telescope reconfigurations by eliminating the need to remove or install the tertiary mirror module. 

We proposed successfully to the National Science Foundation (NSF) Major Research Instrumentation (MRI) Program in 2013 for funding of the K1DM3 project.  We received the full award requested and the total project budget (\$2.1M USD) includes a 30\% cost-share from WMKO, the UC Observatories, and the University of California Santa Cruz.  The project entered its Detailed Design phase in October 2015 and this paper presents the Detailed Design. The K1DM3 device will enable astronomers to swap between any of the foci on Keck 1 in under 2 minutes, both to monitor varying sources (e.g. stars orbiting the Galactic center) and rapidly fading sources (e.g. supernovae, flares, gamma-ray bursts).   The design consists of a passive wiffle tree axial support system and a flexure-rod
lateral support system with a 4.7\,arcminute field-of-view mirror.  The mirror assembly is inserted into the light path with an actuation system and it relies on kinematic couplings for achieving repeatable, precise positioning.   The actuation system may rotate (partially
when retracted or fully when deployed) 
on two bearings mechanized with a pair of drive motors.  
It is our goal to commission K1DM3 at WMKO by March 2017.

\begin{figure}[h!]
\begin{center}
 \vskip -0.1in
 \includegraphics[width=3.0in]{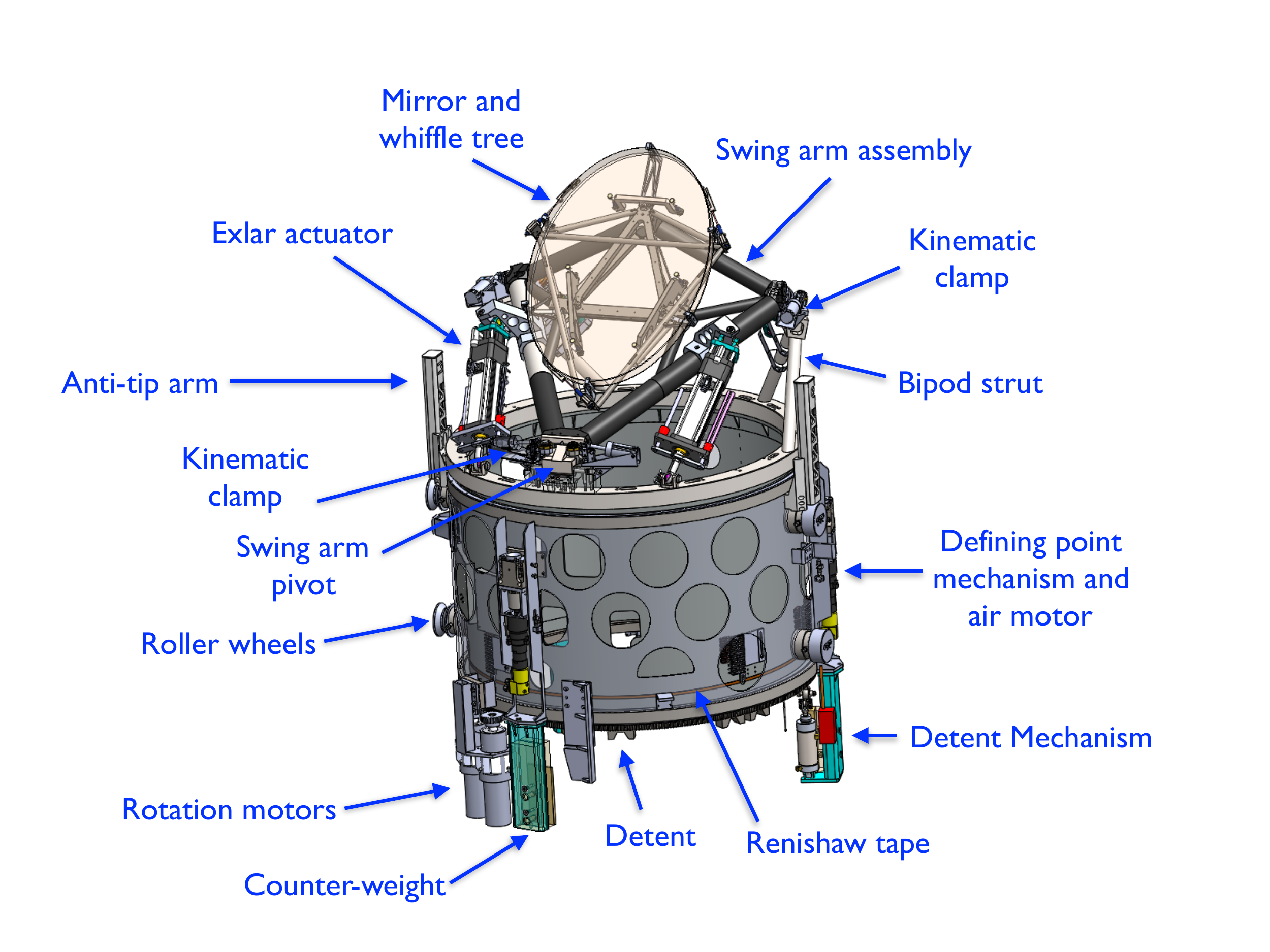}
 \includegraphics[width=3.0in]{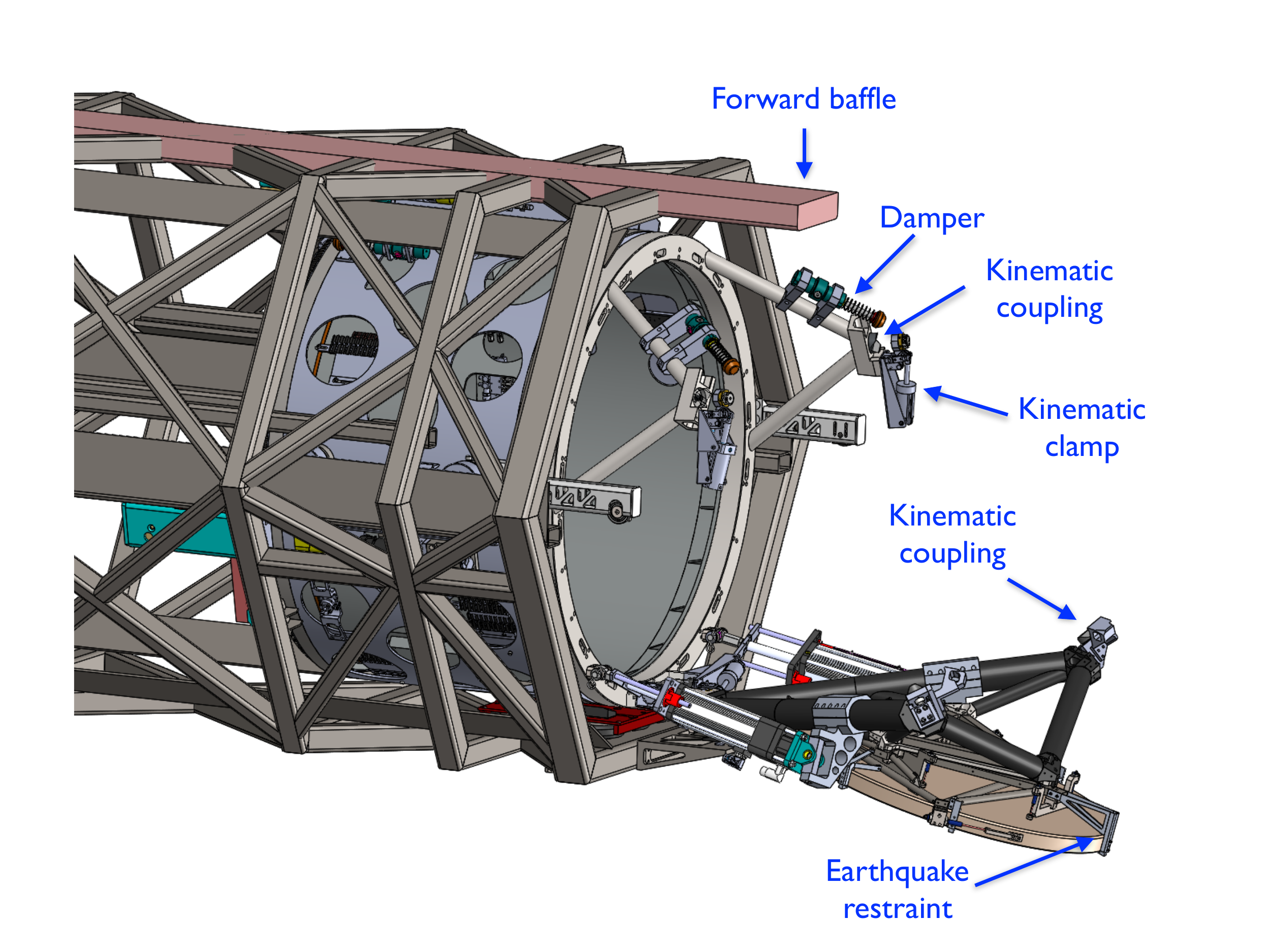}
\vskip -0.10in
 \caption[(left) Schematic of the K1DM3 system. (right)
 K1DM3 module installed in the Keck~I tertiary tower.]{\footnotesize
   (left) Schematic of the K1DM3 system. (right)
   K1DM3 module installed in the Keck~I tertiary tower.
}\label{fig:K1DM3_overview}
\vskip -0.1in
\end{center}
\end{figure}

Figure~\ref{fig:K1DM3_overview} shows the overall configuration of the K1DM3 module and 
the module installed in the tertiary tower. 
The K1DM3 module consists of a light-weighted fixed outer drum and a moveable inner drum.  The inner drum is supported at each end by 4-point contact ball bearings. The lower bearing has a ring gear that is driven with a pinion gear by a servo motor system. An absolute position encoder is used to measure the position of the rotating drum. The tertiary mirror is supported by axial and lateral supports attached to a whiffle tree structure.  This whiffle tree connects the mirror and support structure to a swing arm system.  
In turn, this swing arm moves the mirror between the deployed and retracted positions, driven by two linear actuators. The top of the drum supports the swing arm in the deployed position through a bipod structure with two defining points (at the right side of the figure) and a third defining point at the hinge point of the swing arm (the third defining point is not visible in the figure). The swing arm is locked in the deployed position by 
a set of 4 clamping mechanisms. 
No power is required to maintain the mirror in either the deployed or retracted positions. In order to set the swing arm into
the kinematics, the deployment process will be
performed at the elevation angle where the kinematics are 
oriented normal to gravity (\depang).
We will retract the mirror at specific rotation angles, one of two
positions where power and ethernet is supplied to the swing arm.

Full rotation of the module drum is possible with the mirror deployed. 
Interference with items at the top of the tertiary tower limit the
rotation when K1DM3 is retracted.
When the mirror is deployed, there are six positions used to direct the light to one of the two Nasmyth focal stations or one of the four bent Cassegrain positions. Each of these deployed positions is held by a detent mechanism engaging a v-groove. The detent mechanism is engaged by a pneumatic cylinder and retracted by a spring.

The K1DM3 module is inserted into the tertiary tower from the telescope's Cassegrain platform and moved through the tertiary tower to its operating position on a pair of rails. Guide rollers mounted on the outer drum support the module on the tracks. When the module is installed in the tower it is held in position using three defining point mechanisms equipped with kinematic mounting points that are engaged and disengaged by three air motors.  The kinematic mounting points ensure repeatable positioning.
K1DM3 is designed to be coated with Al at WMKO.

\section{Key Requirements}\label{sec:key}

The K1DM3 team generated a requirements document to help guide the design of our module \cite{k1dm3_requirements}.  Below is a short list of key requirements that have the most impact to our design.  
This paper will focus on the opto-mechanical design;
a brief summary of electronics and software is given in the Appendix.

\subsubsection{Optical Requirements}
\label{sec:opt_req}

\begin{my_enumerate}
\item  The K1DM3 tertiary mirror will be sized to provide an unvignetted 
4.7\, arcminute diameter field of view at the Nasmyth foci. 
\item  The K1DM3 module will not vignette the LRIS or MOSFIRE FOVs when the mirror is fully retracted. The K1DM3 system will not vignette M1 or M2 when the mirror is deployed or retracted.
\item  The surface of the K1DM3 mirror will give an 80\% enclosed energy EE80 in a 0.054\,arcsecond diameter aperture. This corresponds to a surface flatness specification of 9.7E-7 (rms) slope error\cite{k1dm3_mirror_spec}.
\item  The mirror shall be supplied uncoated and shall be coated with bare aluminum by WMKO. 
\end{my_enumerate}

\subsubsection{Mechanical Requirements}
\label{sec:mech_req}

The following requirements primarily concern the motion of the K1DM3 system when installed. 

\begin{my_enumerate}
\item  The mirror will deploy or retract in less than 120 seconds. 
\item  The K1DM3 module shall be provided with a rotator mechanism that serves to point the deployed tertiary mirror at the desired Keck I Nasmyth or bent Cassegrain focal position by rotating the mirror about the telescope optical axis. When the mirror is positioned at one of the six focal station positions it shall be locked in place by a detent or other means. When deployed, the mirror will be able to rotate about the telescope optical axis at a speed of at least 6\,degrees.
per second. 
\item  The K1DM3 module shall not radiate more than 5 watts of heat into the telescope dome ambient environment during an observation. 
\item  The K1DM3 module must not weigh more than 1000 kg. 
\item The structure of the K1DM3 module shall meet the zone 4 earthquake survival requirements of Telcordia Standard GR-63-CORE, 
"NEBSTM Requirements". 
\end{my_enumerate}

\subsubsection{In-beam Positioning Requirements}
\label{sec:positioning}

The current M3 is offset from this ideal position, because of the as-built locations of M1 and M2, the as-built tertiary tower, and/or errors in the original alignment procedure \cite{k1dm3_asbuilt}. Because the Nasmyth instruments on K1 (HIRES, OSIRIS) have been aligned to the existing M1-M2-M3 telescope, we endeavor to replicate (to a specified tolerance) the position of the current M3. 
The following requirements describe performance of the system when deployed relative to the desired location for the K1DM3 mirror. Again, this ``desired location'' may or may not be the optimal position for a perfect telescope system 
(see \cite{k1dm3_positioning} for further details).

Regarding the stability of K1DM3 positioning during an observation (e.g. to vibrations), established convention is to allow uncorrelated effects on image quality at the level of 10\% of the seeing disk. Based on this convention, for 0.4\,arcseconds seeing, translation of the mirror along the telescope X or Z axes should be no more than 29 microns. We adopt this as an rms constraint. Confining the motion to $\pm 29$\,microns (rms) places a stability requirement on tip and tilt of the tertiary of 0.65\,arcseconds and 0.46\,arcseconds (rms).

Synthesizing the above discussion, we derive the following requirements regarding the positioning of the K1DM3 mirror when deployed:

\begin{my_enumerate}
\item  The K1DM3 mirror will position to an accuracy of 725 microns along the telescope X and Z axes (rms).  We adopt the same requirement for repeatability. 
\item  The K1DM3 mirror will position to the nominal rotations of tilt and tip to 11.5\,arcsecond and 16.5\,arcsecond (rms) respectively. We adopt the same requirement for repeatability\cite{k1dm3_positioning}.
\item  The K1DM3 mirror will be held stable to displacements in the telescope X and Z axes to 29 microns (rms).  
\item  The K1DM3 mirror will not move in tip and tilt due to external influences (vibration) by more than 0.65\,arcseconds and 0.46\,arcseconds (rms) respectively. 
\end{my_enumerate}

\subsubsection{Interface Requirements}
\label{sec:interface_req}

The following list of requirements relate to the interface between the K1DM3 system and the K1 telescope.

\begin{my_enumerate}
\item  The K1DM3 module shall be designed for installation in the Keck I tertiary tower using the same defining points provided for the existing Keck I tertiary mirror module. All adjustments to align the K1DM3 module in the telescope shall be made by adjusting the defining point halves located on the K1DM3 module. 
\item  The K1DM3 module shall be compatible with the existing module insertion and removal rails
provided in the Keck I tertiary tower. 
\item  The K1DM3 tertiary mirror shall be removable for recoating and shall be provided with an adapter as required to permit the use of the existing Keck I tertiary mirror handling fixture when the mirror is removed for recoating. 
\end{my_enumerate}


\section{Optical Design}\label{sec:optical}

\subsection{Mirror Design}
\label{sec:mirror}

The K1DM3 system will provide a new tertiary mirror for the Nasmyth and bent-Cassegrain foci of the Keck I telescope.  
The flat mirror is made of 
Zerodur glass and is shaped as an ellipse with major 
axis 2a = 901.1 mm and minor axis 2b = 643.0 mm, and 
a thickness of 44.5 mm. 
It has an approximate mass of 51.74\,kg.
The K1DM3 project obtained the mirror blank from the TMT project,
on permanent loan, and a second blank as a spare.

\begin{figure}[h!]
\begin{center}
 \includegraphics[width=2.in]{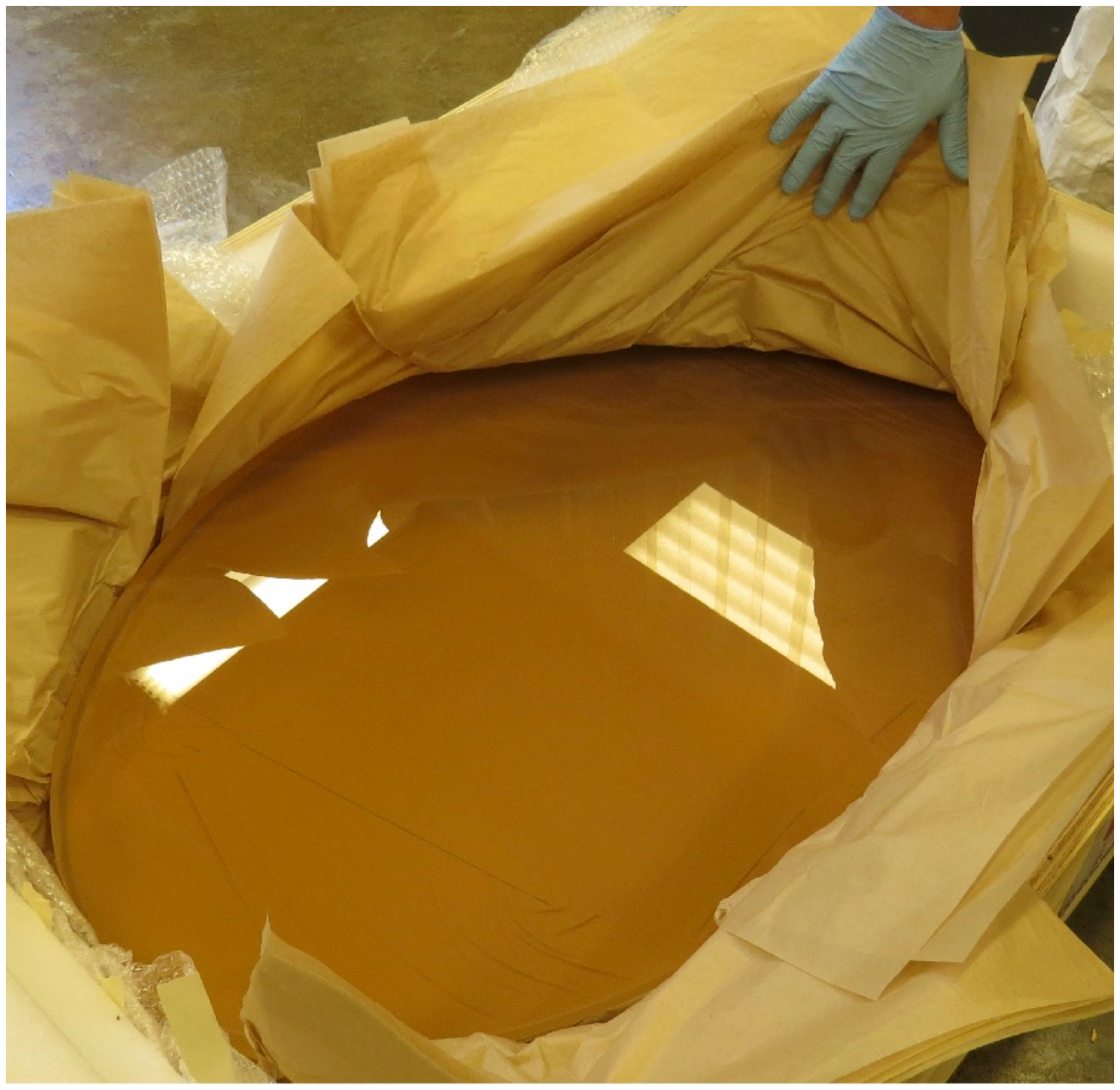}
 \includegraphics[width=3in]{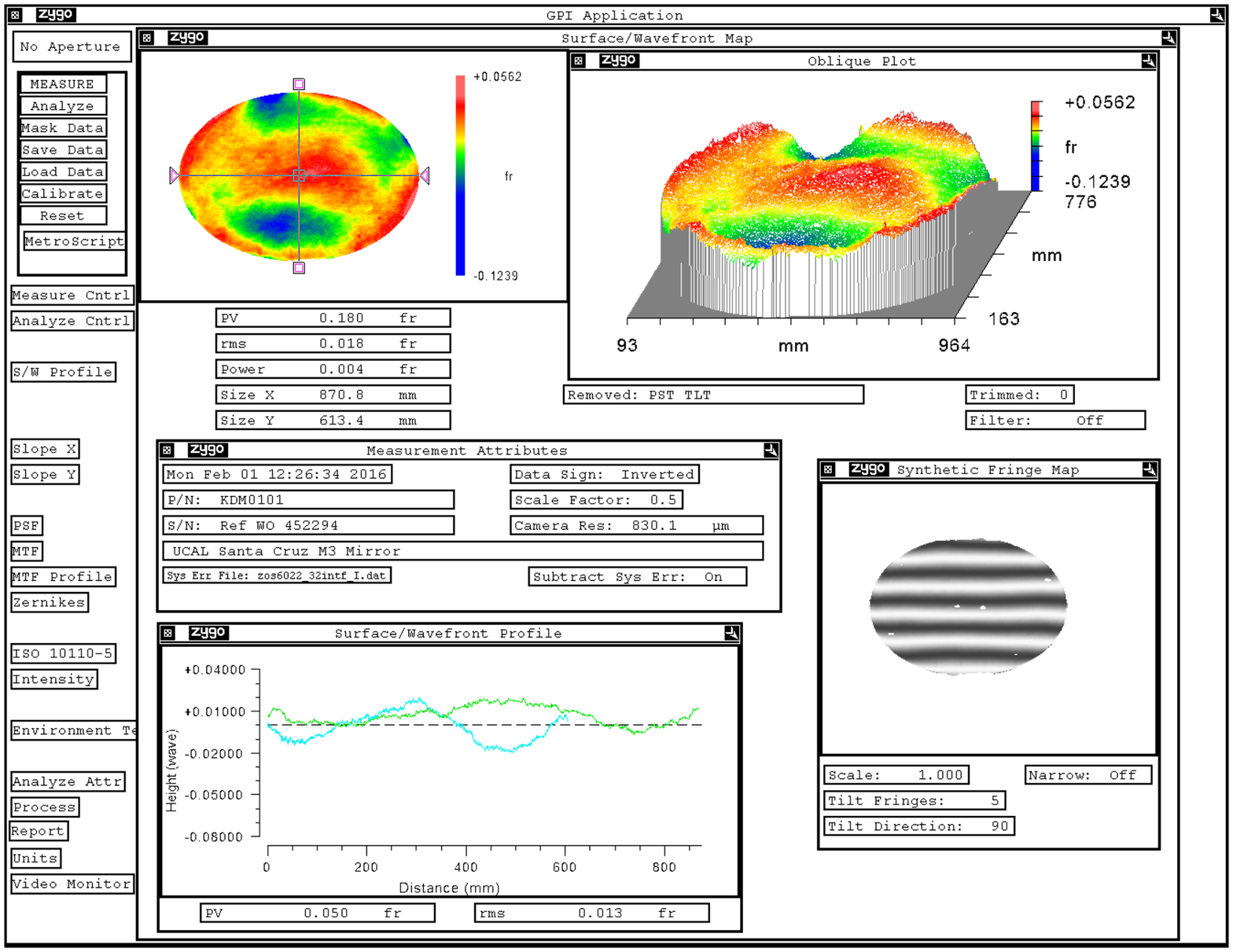}
 \caption[(left) As-built K1DM3 mirror at UCO and (right)
  surface quality measurements on the clear aperture of the 
 as-built mirror for K1DM3.]{\footnotesize
  (left)  Image of the as-built K1DM3 mirror, cut and polished
  by Zygo and now delivered to UCO.
  (right) Surface quality measurements on the clear aperture of the 
 as-built mirror for K1DM3.
}\label{fig:mirror}
\vskip -0.1in
\end{center}
\end{figure}

The mirror was fabricated by Zygo and has been delivered
to UCO (Figure~\ref{fig:mirror}).  The figured mirror has
an edge exclusion
of approximately 15\,mm width around the outer circumference
that has poorer image quality.  This reduces the clear
aperture FOV to approximately \fov\ (see \cite{k1dm3_zygo} for 
further details).
Within the clear aperture,
the mirror was polished by Zygo to 0.2\,nm (rms) surface roughness 
(Figure~\ref{fig:mirror})
and to meet a (60-40) scratch/dig surface quality per MIL-PRF-13830B.  The non-optical surface finish is R2 ground flat, 400 grit finish or better.  The reflective surface was polished to a P-V of 57\,nm and 
a surface error of 6\,nm. 
The mirror will be delivered uncoated and later coated with bare Aluminum using the coating chamber at WMKO. 
Multi-layered-protected silver coatings were considered by the
project but deemed too expensive given the associated
risks with transport and handling.

\subsection{Vignetting of the Cassegrain Instruments when Retracted}
\label{sec:vignette}

\subsubsection{Design Description}
A key aspect of the K1DM3 system is to enable observations with the mounted Cassegrain instrument by retracting the tertiary mirror out of the beam on demand.  This is a unique functionality in comparison to the existing tertiary module.  We have designed K1DM3 accordingly and have also considered carefully the dimensions and positions of the module and retracted mirror to avoid vignetting the light arriving at the Cassegrain focus.  We summarize the main issues that have been addressed and refer 
to \cite{k1dm3_positioning} for further details.

\begin{figure}[h!]
\begin{center}
\includegraphics[width=4.0in]{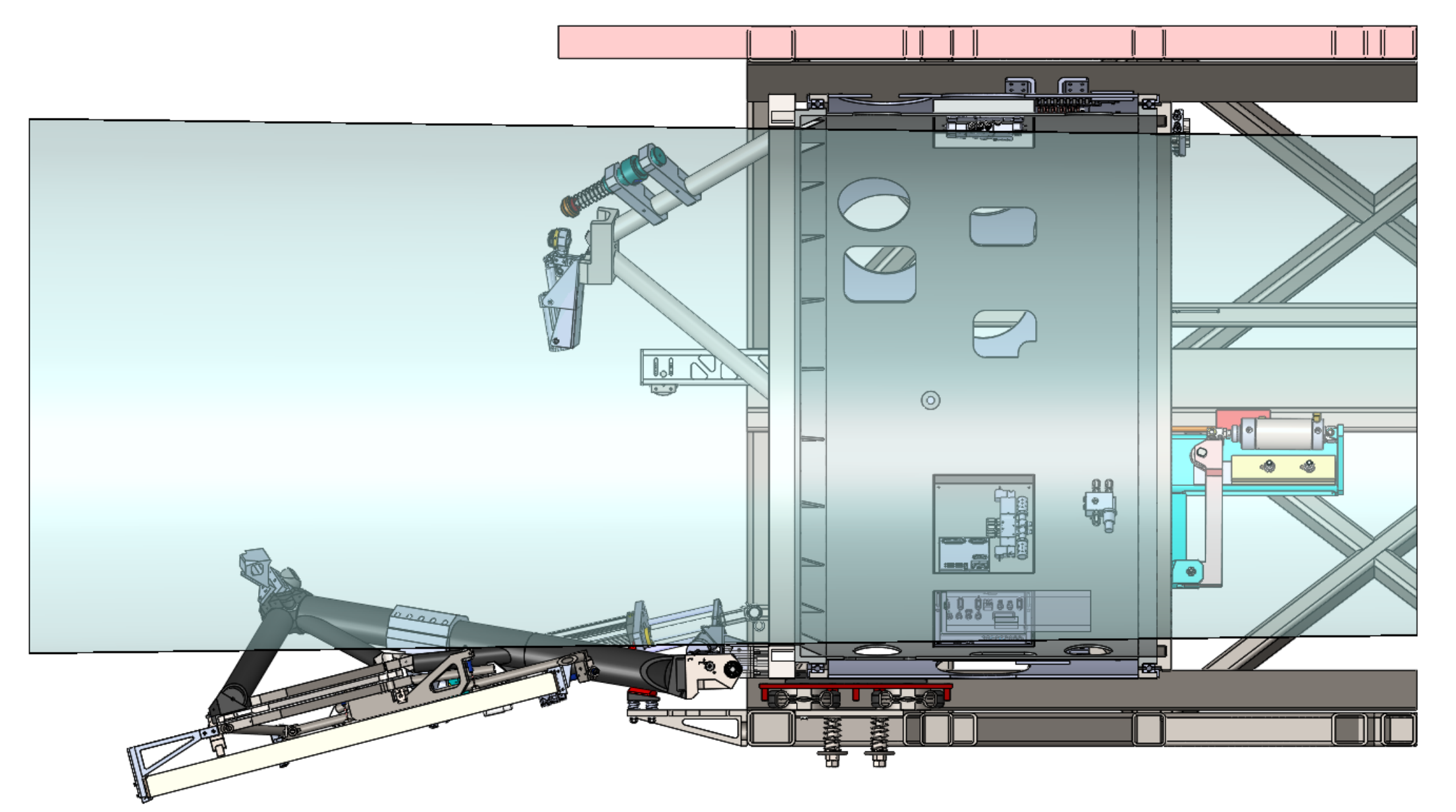}
\includegraphics[width=2.0in]{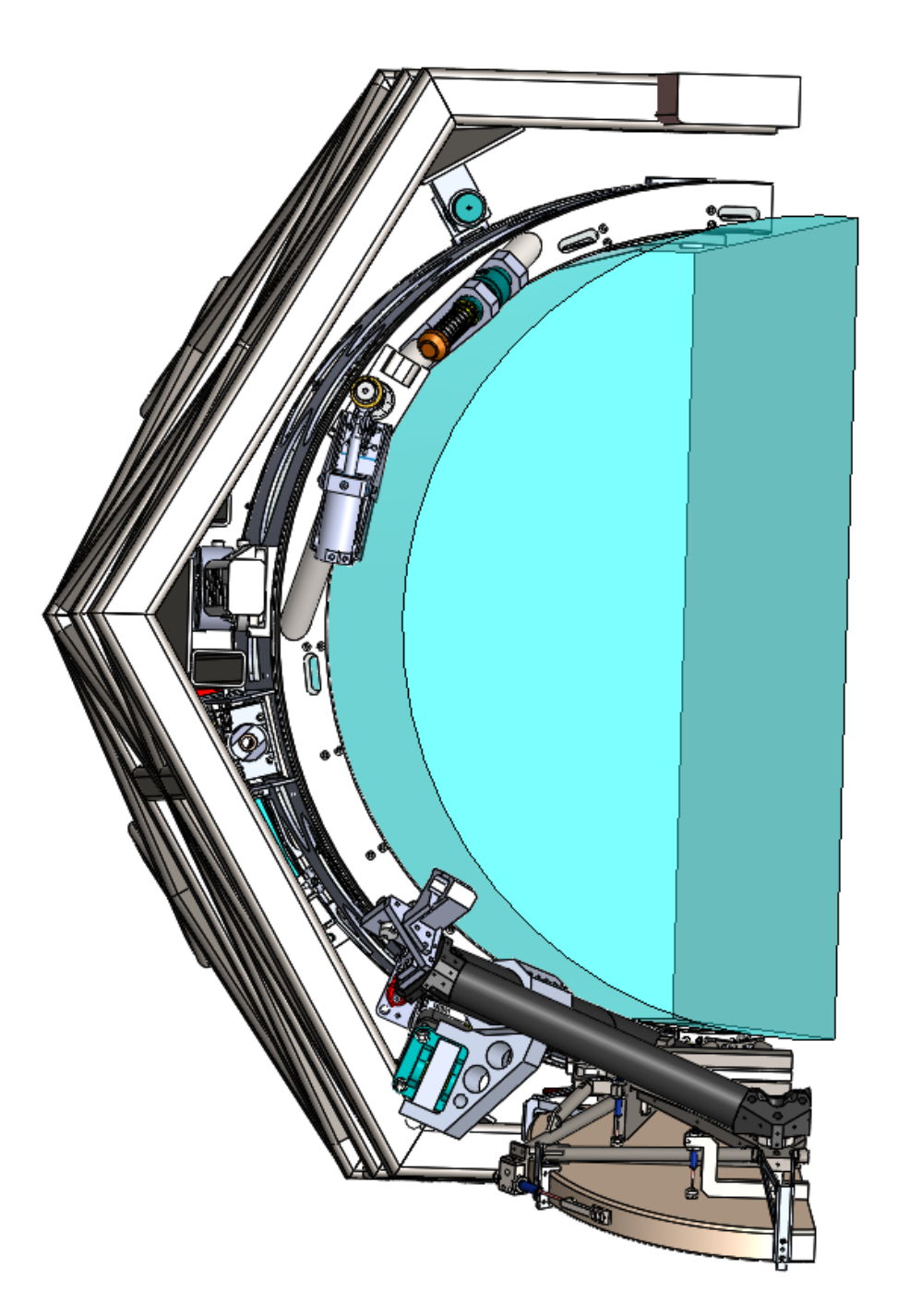}
 \caption[Two views of K1DM3 in the retracted position]{\footnotesize
 Two views of 
K1DM3 in the retracted position showing proximity of the M2 to Cassegrain beam footprint
(colored shroud).
}\label{fig:retract}
\vskip -0.1in
\end{center}
\end{figure}

We face significant challenges when M3 is retracted to avoid vignetting the rays from M1 to M2 and, at the same time, avoid vignetting the rays from M2 to the Cassegrain instruments. 
An additional concern is vignetting related to the LRIS ADC
(see \cite{k1dm3_ADC} for further details).
When M3 is retracted, it will be held above the module and the tertiary tower with the reflective surface facing away from the optical axis as shown on the left side of Figure~\ref{fig:retract}). 
In this position, we must avoid the rays travelling to M1 and (more importantly) the converging rays from M1 to M2. We will retract the center of M3 to this position: a height of 267.16\,mm above the elevation axis and radially offset by 759\,mm from the optical axis, and at an angle $\alpha=104.5$\,degrees (where $\alpha=45$\,degrees is the deployed position and $\alpha=90$\,degrees is parallel to the optical axis). The result is no vignetting of the converging rays from M1 to M2 over the full 20\,arcminute Cassegrain FOV. 

Presently, there are two Cassegrain instruments commissioned on Keck I (with none additional currently planned): LRIS with a 6\,arcminute x 8\,arcminute FOV located 7\,arcminute off-axis and MOSFIRE with an on-axis FOV of 6.14\,arcminute x 6.14\,arcminute. Each instrument has an off-axis guide camera. The rectangular fields of view of the science and guide cameras for LRIS and MOSFIRE generate rounded ``footprints'' 
normal to the optical axis that one must avoid to prevent vignetting. 
The dimensions and shape of these footprints decrease as the beam converges from M2 to the Cassegrain focus (i.e. as a function of elevation along the optical axis).

\subsubsection{Design Analysis}
Analysis of vignetting of the retracted K1DM3 mirror on rays traveling from M1 to M2 was performed with the Zemax software package. We implemented the user-defined aperture (UDA) for the Keck primary mirror, a circular M2 mirror with radius of 700 mm, and the apertures needed to represent the M2 spider. In addition, we modeled obscuration by the tertiary tower as a hexagon with sides of 880.4 mm placed at a height of 3451.6 mm above the primary. We also modeled the obscuration by the secondary structure as a hexagon with sides of 1.32 m. Lastly, we calculated the vignetting of rays by the retracted M3 as an elliptical aperture held at the position defined above. The secondary mounting structure casts a sizable shadow on the surface of M1. The retracted K1DM3 fits entirely within this shadow. The rays converging from M1 to M2 present a tighter constraint, but we find that K1DM3 does not vignette any of these rays if the angle of the retracted mirror is less than approximately 105\,degrees.
The footprints described above were generated with an IDL code using simple geometrical arguments and the known dimensions of the K1 telescope. We then generated UDAs at several heights above the primary and imported these within the as- built Zemax models for the LRIS and MOSFIRE designs. We verified with Zemax that these footprints are correctly sized.

\begin{figure}[h!]
\begin{center}
 \includegraphics[width=5in]{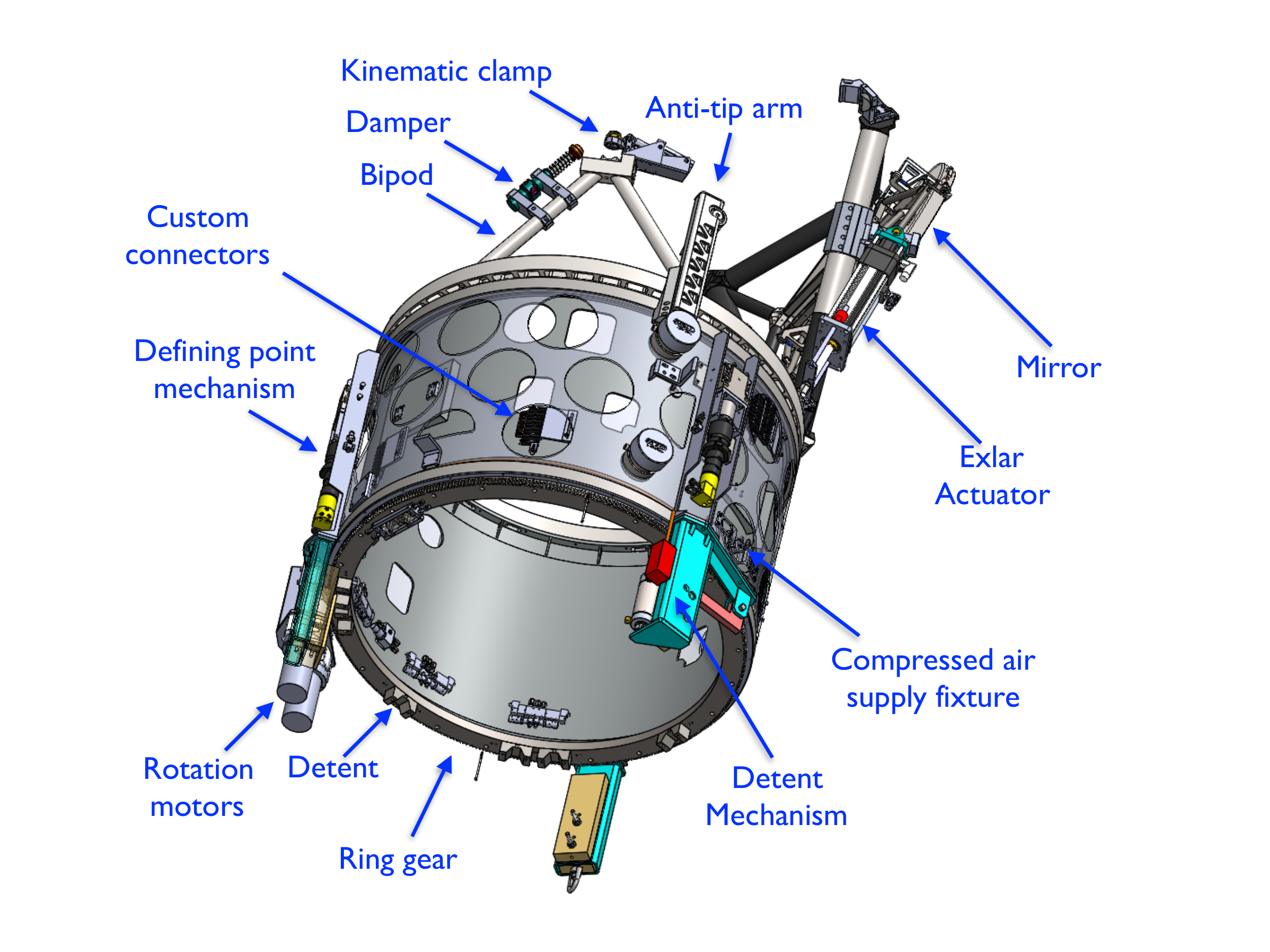}
 \caption[K1DM3 module as viewed from below]{\footnotesize
K1DM3 module as viewed from below, with callouts.  See
Figure~\ref{fig:K1DM3_overview} for a view from above.
}\label{fig:module_bottom_view}
\vskip -0.1in
\end{center}
\end{figure}

\section{Mechanical Design}

\subsection{Mirror Assembly}\label{sec:assembly}



\subsubsection{Design Description}
The mirror for K1DM3 requires a support structure that will (i) maintain the mirror's figure under varying gravity vectors and temperature changes; (ii) interface the mirror with the deployment (swing arm)
mechanism; (iii) ensure the safety of the system during an earthquake; and (iv) provide a means to coat the mirror within the WMKO coating chamber.

\begin{figure}[h!]
\begin{center}
\includegraphics[width=2.5in]{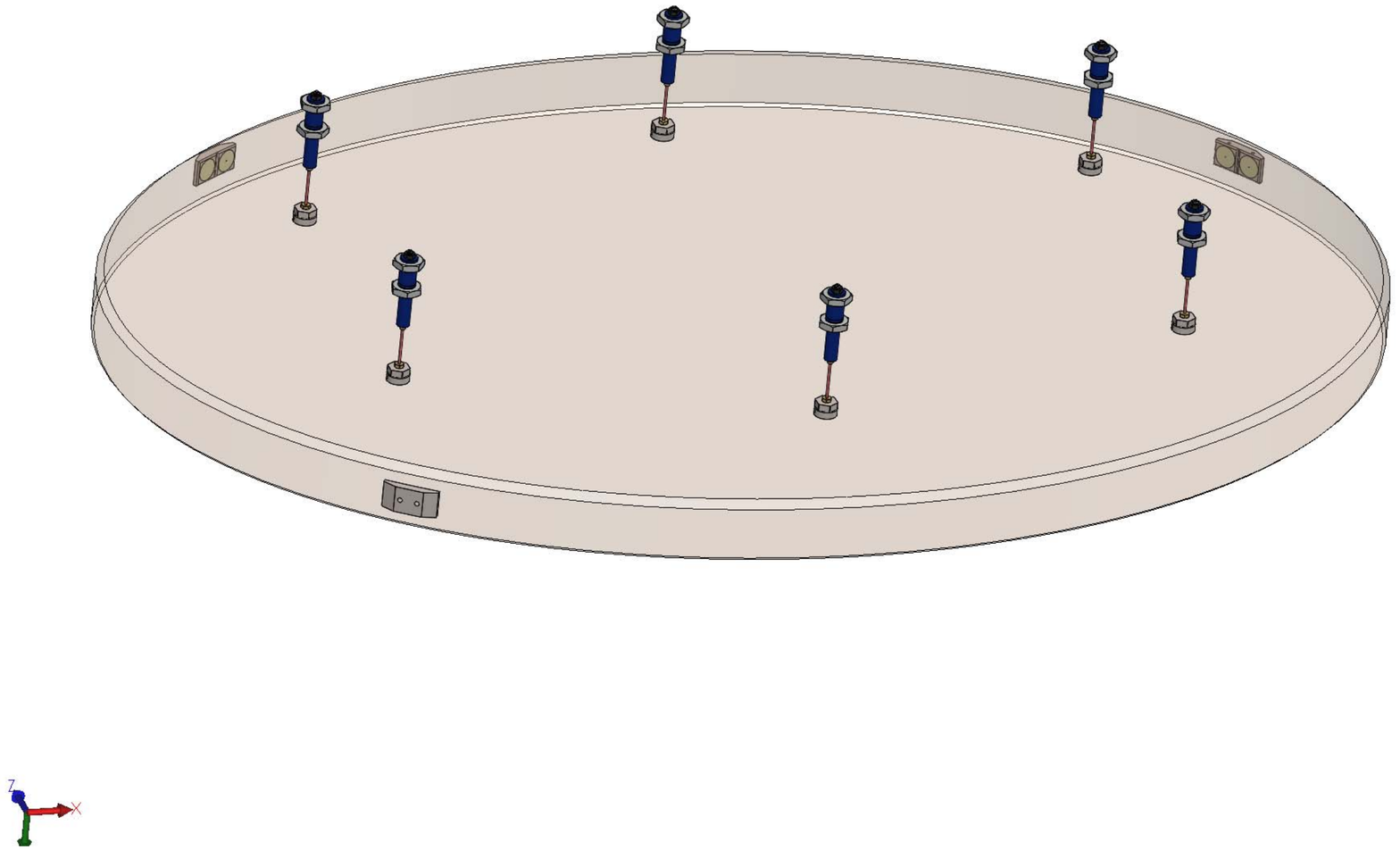}
\includegraphics[width=2.5in]{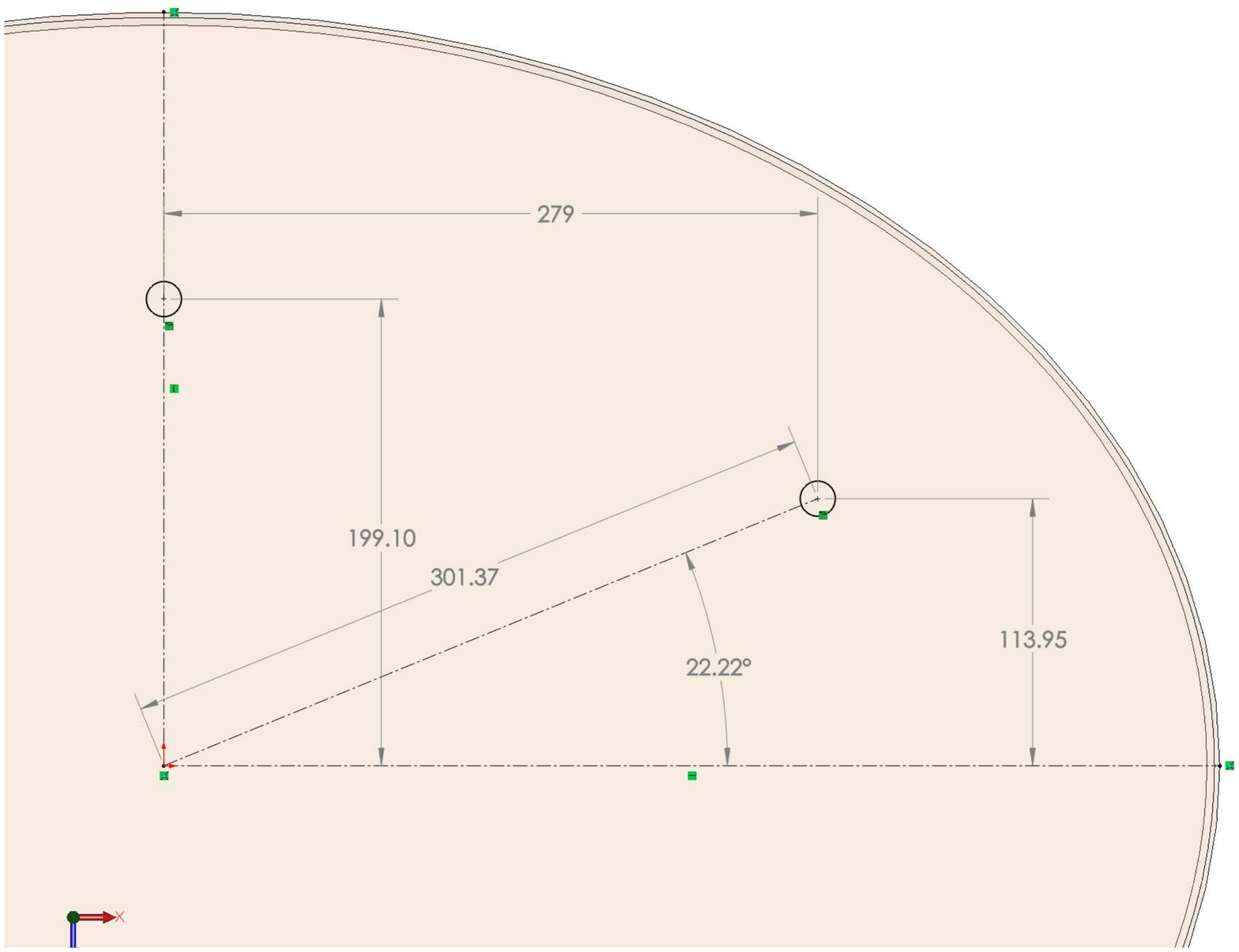}
 \caption[Axial rod placement]{\footnotesize
The left CAD image shows the six axial support rods attached to the back of the K1DM3 mirror. The diagram on right shows the placement of two of the rods indicated by open circles. The measurements are referenced from the major and minor semi-axes of the mirror.
}\label{fig:axial_support}
\vskip -0.1in
\end{center}
\end{figure}

For axial support, the K1DM3 design uses six rods inserted into pucks glued to the back (i.e. non-reflective) side of the mirror. These rods are 1.7\,mm in diameter, have 60\,mm free length, and will be made of AISI M2 steel. The pucks are Invar and will be glued with an epoxy adhesive. The axial rods are screwed into the pucks. The layout of these six axial support rods is shown in 
Figure~\ref{fig:axial_support}. 
The positions for the rods were determined from finite element analysis (FEA) to minimize the deflections of the mirror normal to its surface.
These were updated to reflect the final dimensions of the mirror.

\begin{figure}[h!]
\begin{center}
\includegraphics[width=2.5in]{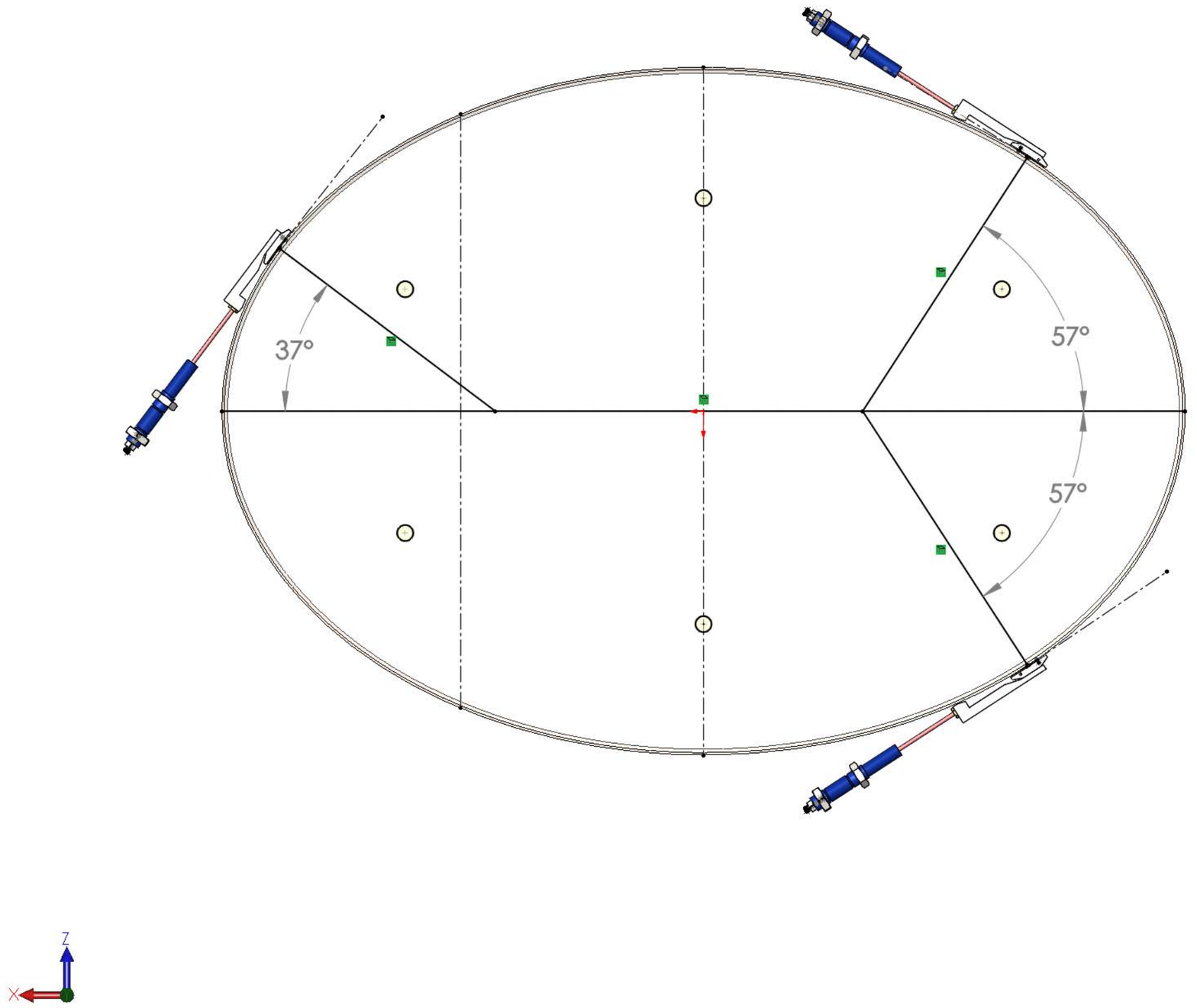}
\includegraphics[width=2.5in]{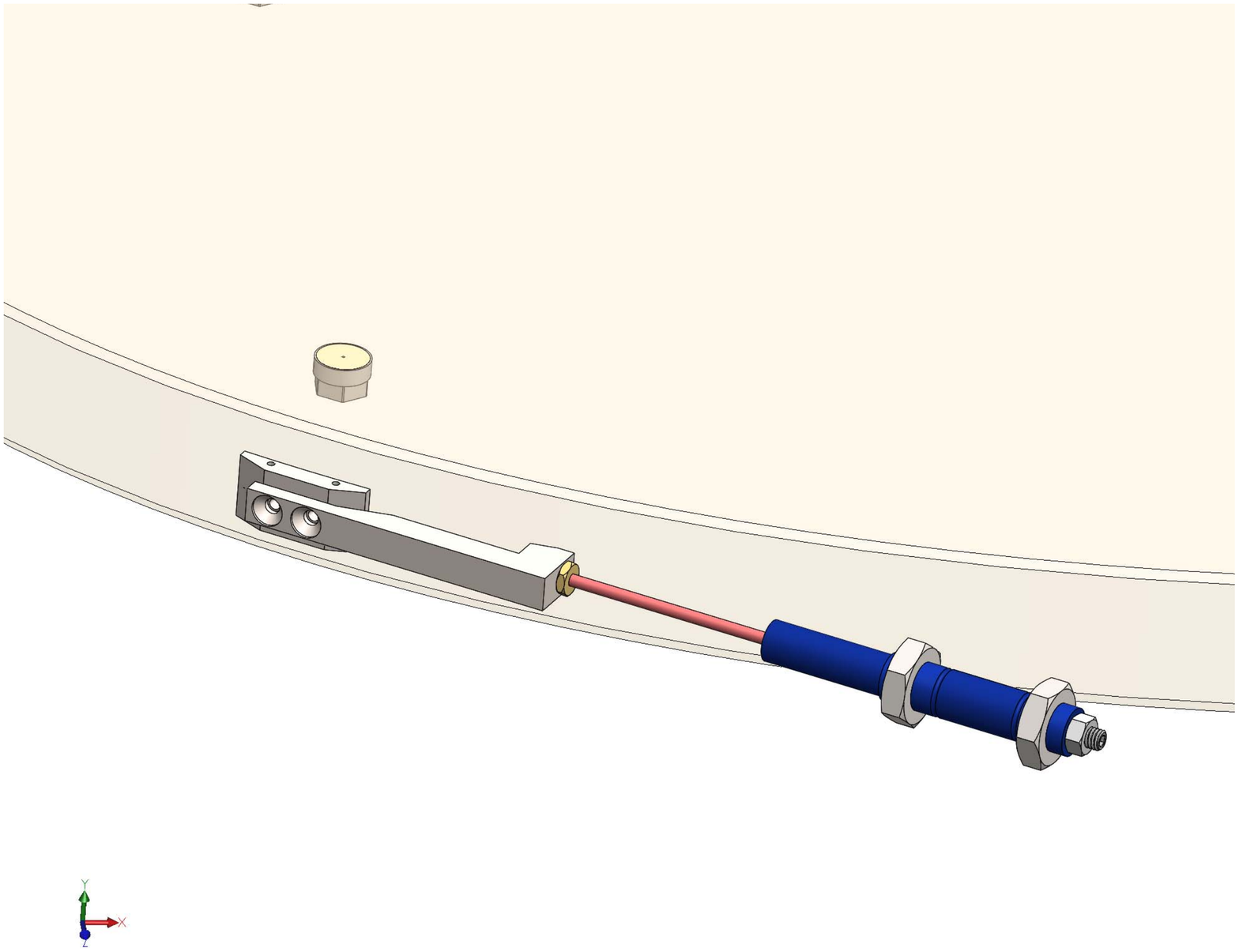}
 \caption[Lateral rod placement]{\footnotesize
 Left diagram shows the placement of the three lateral supports along the outside edge of the mirror. Right image is a close-up view of one of the lateral supports screwed to a puck that is glued to the edge of the mirror.
}\label{fig:lateral_support}
\vskip -0.1in
\end{center}
\end{figure}

Lateral support is provided by three flexure rods fastened
to pucks glued to the mirror's edge, similar to the axial flexure rods. These are at approximately the major axis of one side and two other positions opposite (Figure~\ref{fig:lateral_support}). 
These rods are 125\,mm free-length and 3.0\,mm in diameter, and are AISI
M2 steel. 
The pucks were designed to ensure a less than
15\,micron difference in the glue thickness across the surface.
For expected temperature changes,
the design results in 130\,PSI (static) and 655\,PSI (dynamic)
of stress in the bonds and glass.
These stresses are significantly smaller than the estimated 
3000\,PSI of stress that the bonds can endure and are also
significantly smaller than the limits that we wish to maintain
for Zerodur.

\begin{figure}[h!]
\begin{center}
\includegraphics[width=2.5in]{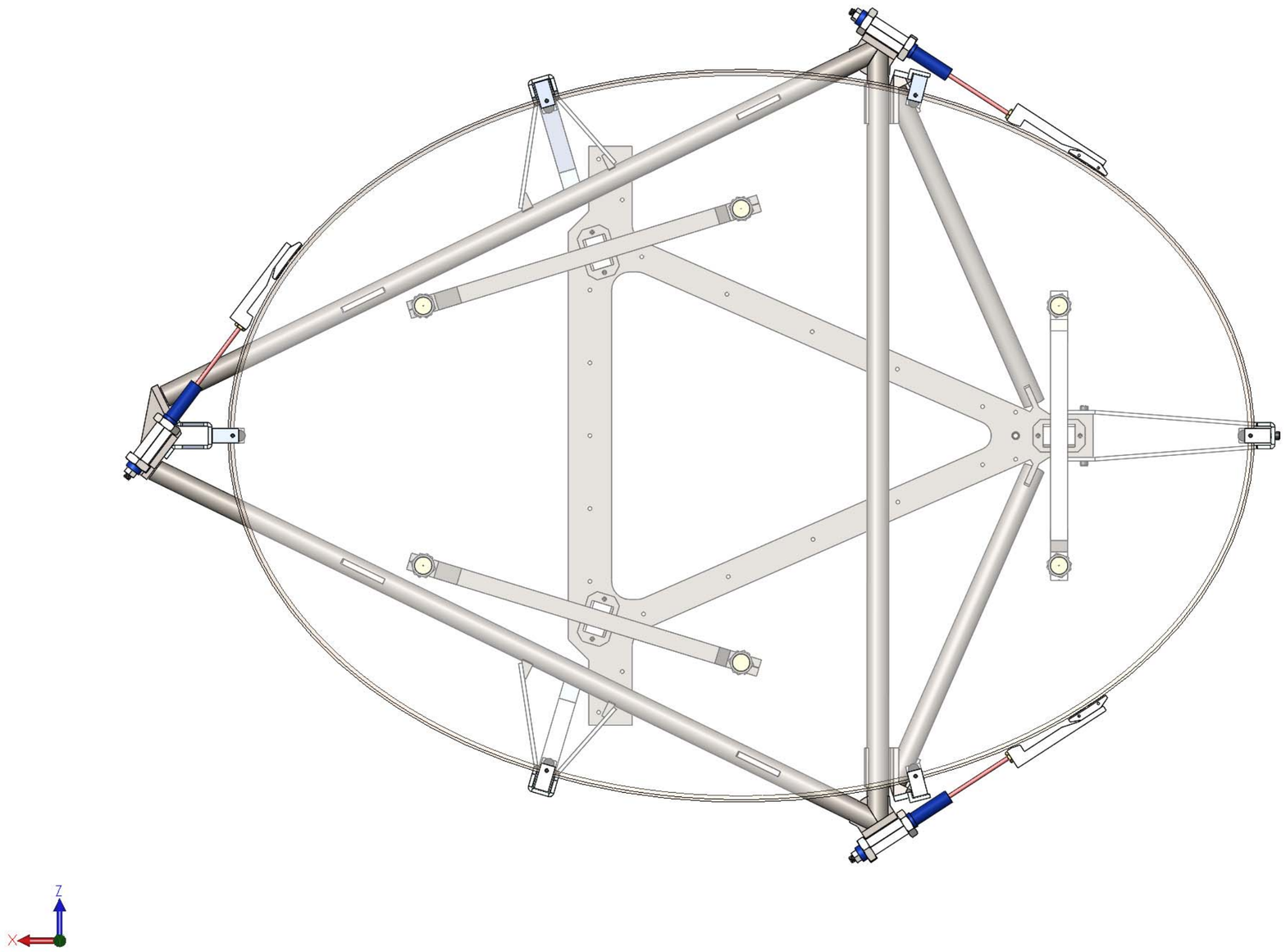}
\includegraphics[width=2.5in]{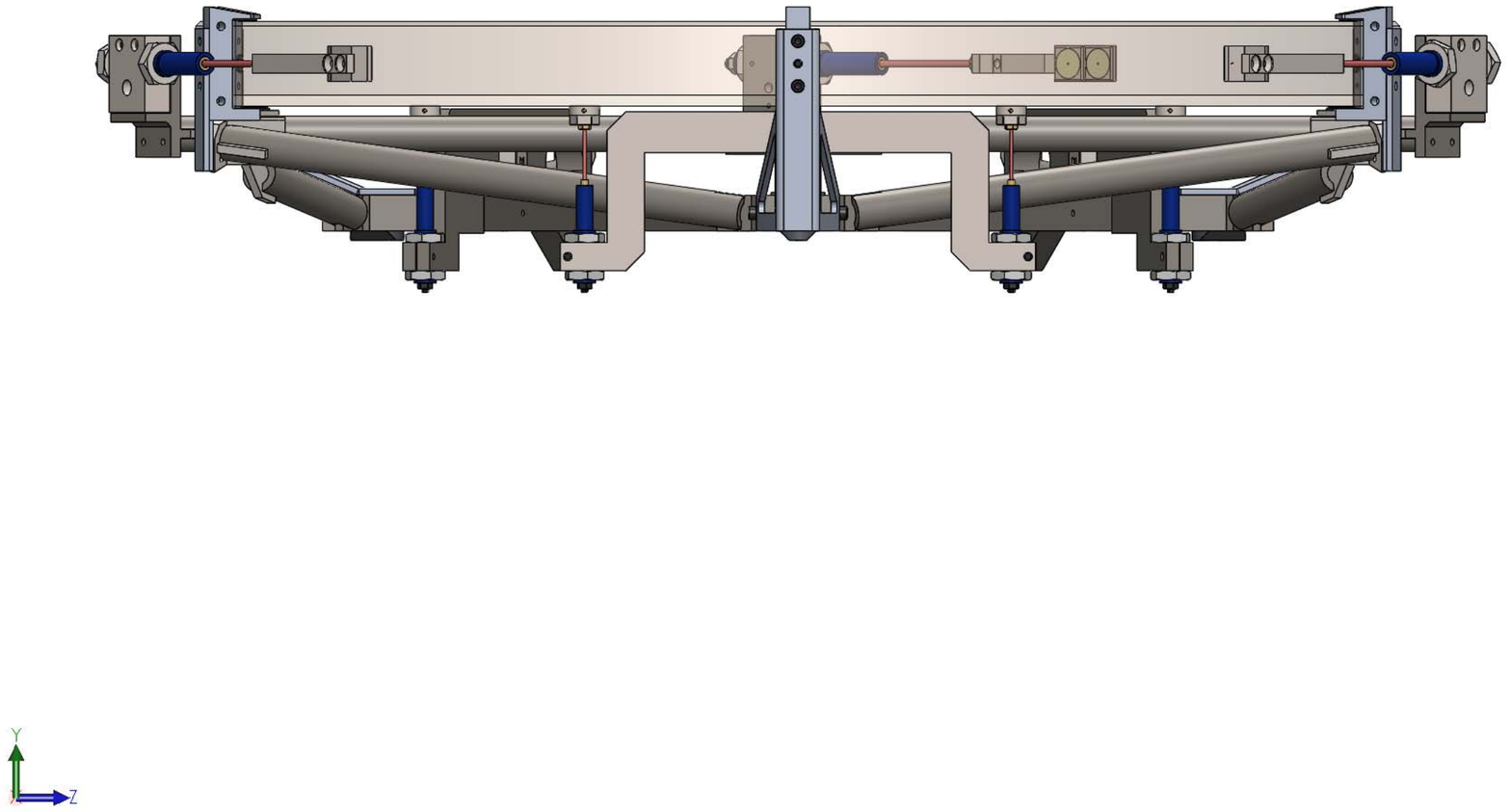}
 \caption[Whiffle-tree support structure]{\footnotesize 
 Top and side views of the whiffle-tree support structure for K1DM3.
}\label{fig:whiffle}
\vskip -0.1in
\end{center}
\end{figure}

The axial and lateral rods are integrated within a whiffle-tree support system, shown in Figure~\ref{fig:whiffle}. The whiffle tree uses 
0.75\,inch diameter struts (5/32\,inches thick)
in a determinate truss pattern. It allows the Mirror Assembly to be bolted to the Deployment System and then removed for re-coating. 
The total mass of the Mirror Assembly including the mirror 
is estimated to be 70\,kg.

The mirror assembly includes three kinematic fixtures
(0.5\,meter radius spheres in v-grooves) 
that interface it with the swing arm assembly.
This allows one to remove the mirror assembly
from the K1DM3 system
for coating.  It then enables one to repeatably and precisely
reattach the mirror assembly to the swing arm 
assembly.  The design also allows
for fine adjustment of the kinematics during the alignment 
phase.

We have designed an earthquake restraint system consisting
of 6 clips affixed to the mirror and the whiffle tree
truss system. 
These are spaced approximately
evenly around the circumference (Figure~\ref{fig:whiffle}).  
They are made of Aluminum with Teflon pads positioned
in normal operations with a gap (i.e.\ not touching the
mirror).  These are
designed to take the static load 
of the mirror assembly
in the event that the glue bonds fail during an earthquake. 

\subsubsection{Design Analysis}

The guiding philosophy for the Mirror Assembly design is to provide adequate support while minimizing complexity. We started from the current M3 support-structure which utilizes 24 axial rods glued into holes that were drilled into the back surface of the glass. There is one lateral support assembly with pads along a ring that were glued to a hole drilled deep into the glass.

Our first efforts reduced the previous M3 support structure design to six axial rods and lateral support assembly similar to that of the current M3. We believe this satisfied all requirements related to mirror stability. 
We were then encouraged to consider a design without holes drilled into the glass. With additional investigation, we determined that a six axial rod and 3 lateral rod design provides adequate support. The rod assemblies will be attached to pucks glued to the glass.

\noindent
\underline{FEA Modeling:}
The positioning of the axial rods was optimized iteratively in a series of FEA models processed with ANSYS. The modeling and static deflection analysis was performed with traditional 3D, 20-node brick elements which yield displacements (3 degrees of freedom) at the nodal locations. To obtain slopes (rotations) and reasonable statistics, the surface deformations of the top surface were mapped to a denser and uniform shell model. Results of this second model provided the surface slopes (and deflections) over a uniformly distributed surface. These results were exported to Excel for easy processing to obtain statistical values (max, min, rms, etc.). 
Our primary metric in evaluating a given model was the peak-to-valley (PV) deflections over the surface. We examined these with three orthogonal gravity vectors: one normal to the mirror surface and the other two along the directions of the major and minor axes.
We implemented a 
mesh geometry for an ANSYS model of the current design. The model 
assumes the mirror properties described in 
$\S$~\ref{sec:mirror}.

\begin{figure}[h!]
\begin{center}
\includegraphics[width=2.5in]{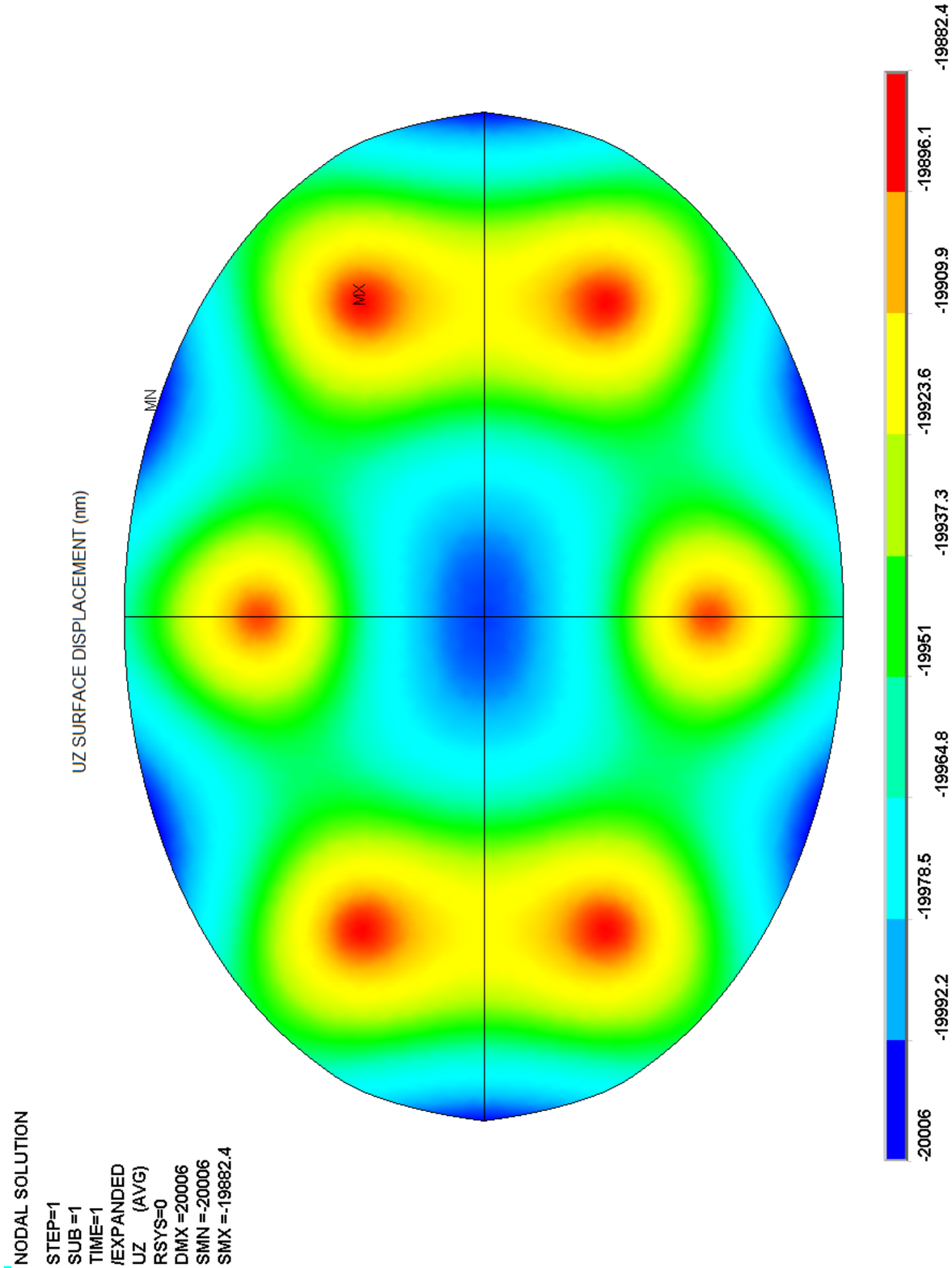}
\includegraphics[width=2in]{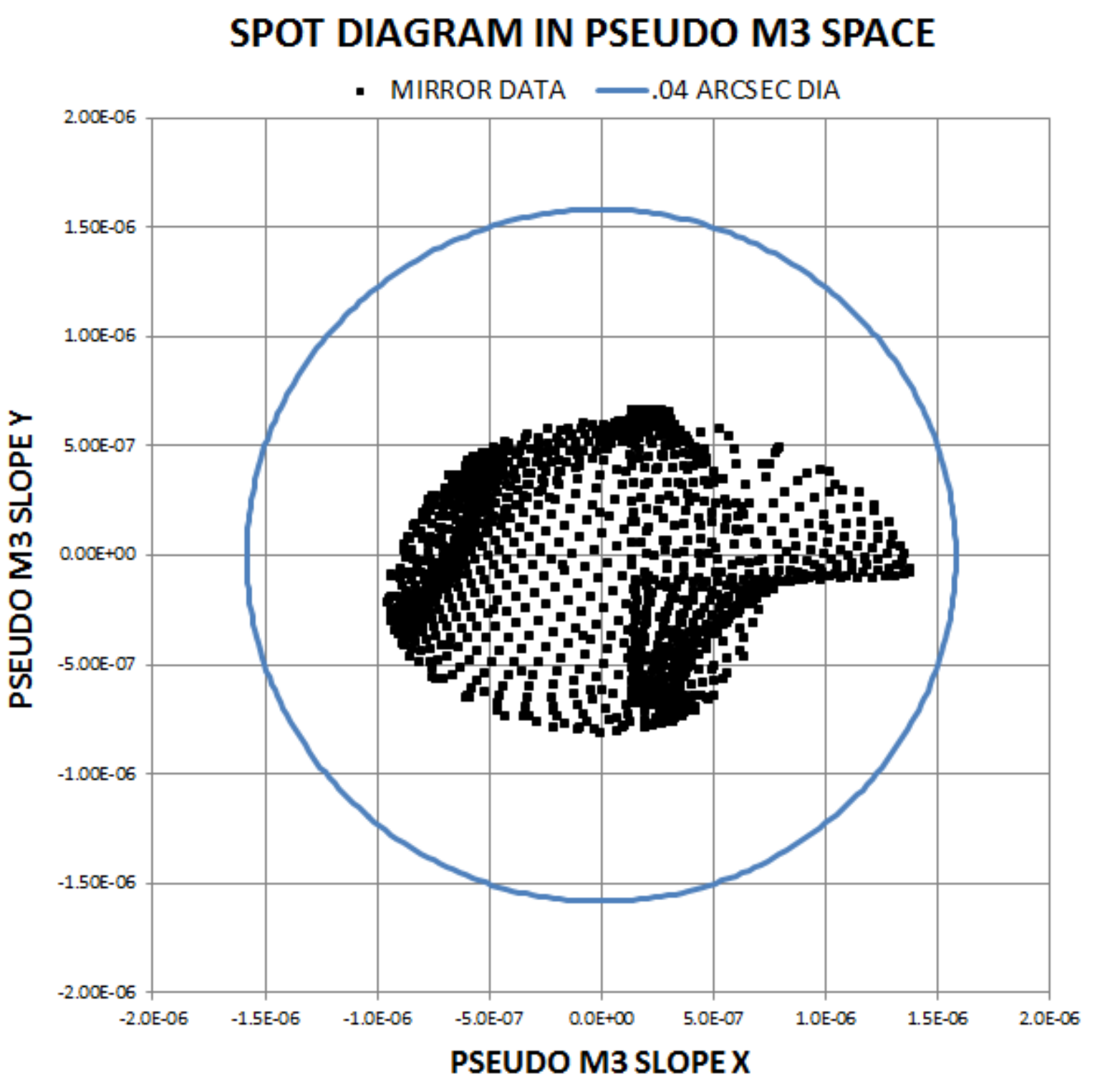}
 \caption[(left) Surface deformation map; (right) spot diagram]{\footnotesize 
 (left)
Surface deformation map for the six point support system due to gravity normal to the mirror. The peak to valley range is 124\,nm. Rms deflection of the entire surface is 39\,nm. 
(right)
 Spot diagram representing where light reflected by the mirror would strike the focal plane. The image is a family of points which are calculated based on the slope error throughout the mirror. 
 The blue circle for reference is 0.04\,arcseconds in diameter. 
}\label{fig:deform}
\vskip -0.1in
\end{center}
\end{figure}

Figure~\ref{fig:deform} is a deflection contour map of displacement (nm) normal to the surface. The load is gravity normal to the mirror surface. This is the most severe loading condition encountered by the mirror during normal operating conditions. The peak-to-valley displacement is approximately 124\,nm with an rms of 29\,nm.
A pseudo spot diagram shown in Figure~\ref{fig:deform} is an assessment of the worst-case deformations reported above. The basis for this diagram is the deformation response of the mirror due to normal gravity. Surface deflections of the model (Figure~\ref{fig:deform}) are mapped to a uniform shell mesh.
The plot is an aggregate of all the points on the uniform mesh. Each dot on the graph represents the two out of plane rotations of the point. X Slope is about the mirror major axis; Y Slope is about the minor axis. For a perfectly flat mirror all points would be at the center. This shows that the rotations are small and the overall image blur is negligible.

\noindent
\underline{CTE:}
Within the K1 dome one may experience changes in temperature ranging from $-10^\circ$ C to 20$^\circ$ C. Thermal expansion of the glass and the mirror support structure will lead to deformations in the mirror surface and its position. The design minimizes the effects of CTE (for a temperature change of 30$^\circ$ C) to about 2.4 nm peak-to-valley, 0.6\,nm rms, 
and a maximum slope error of 0.0011\,arcseconds.

\noindent
\underline{Deformation Summary:}
The error budget for the mirror considered a range of conditions,
as detailed below.
We allow a total rms blur (due to slopes) in the focal plane of 7.3E-8,
$\theta_y$ (slope error about major axis) must be less than 8.42E-7 at tertiary 
and $\theta_x$ (slope error about minor axis) must be less than 1.19E-6.

Rotations refer to the slope in the mirror surface due to deformation. The formula becomes:

\begin{equation}
Maximum\; permissible \; rotation \; = \; \sqrt{\theta_x^2 + 2 * \theta_y^2}
= 1.68 \sci{-6} \; {\rm radians}
\end{equation}

We studied the
the rms slope error for various conditions 
and allowances: (i) the support design error, based on the static response due to gravity normal to the mirror surface, which is the worst case gravity vector; (ii) CTE axial error, which refers to differentials in thermal expansion or contraction which would influence the mirror through adhesive and the Invar pucks attached on the mirror's rear surface. 
(iii) CTE Lateral error, which is the same effect caused by the three attached pucks on the outer edge of the mirror.
(iv) fab errors axial, which refer to a 1N load in plane load caused by the attachment of the axial support system. 
(v) fab errors lateral which is a similar assessment for the lateral support system. 
(vi) moment fab errors which are based on a 0.2 N-m moment error in the respective support systems.
and 
(vii) axial pivot error, which considered a pivot alignment, or location, error of 1 mm. Such an imbalance would result in improper reactions at the six support locations, thereby increasing surface deformations. This is based on worst case gravity acting normal to the mirror surface.

The quadrature
sum of these error terms is $7.14 \times 10^{-7}$
and is well below the maximum allowed value.  

\noindent
\underline{Kinematic Fixtures:}
We will use 3 sphere-in-groove kinematics
(0.5\,m radius spheres) 
to repeatedly and precisely attach the mirror assembly to
the swing arm assembly.  These have a purported repeatability
in position of less than 1\,micron and should contribute 
negligibly to misalignment.  These couplings also have fine
adjustments which will be set during alignment tests at UCO
(i.e.\ prior to delivery to WMKO).

\noindent
\underline{Gluing:}
A critical aspect of the mirror assembly is the gluing of the lateral and axial support pucks to the glass.  The project has chosen to rely on the extensive expertise of WMKO for this process.  Their staff has performed extensive research \cite{kotn_579, kotn_580, kotn_803, kotn_804, kotn_819, kotn_823, kotn_824}
into adhesive selection, application, and strength performance testing which resulted in their choice of the preferred gluing agent (Hysol Loctite E-120HP), finding that it performed well in strength tests and is substantially easier to mix and apply.  It will be necessary to etch the glass with a solution produced by Schott prior to gluing, which follows a rigorous cleaning and surface preparation regimen.

WMKO has invested heavily and developed equipment, material, and procedures for bonding Invar to Zerodur in preparation for their segment repair program.  With their guidance and consultation we will build proof test samples to validate the design to the anticipated loads, including all fixturing, tooling, and part fabrication for the final gluing assembly.

\noindent
\underline{Earthquake restraints:}
We designed the earthquake restraint system to bear the
full weight of the mirror in the event that an earthquake
tears the glue bonds on the axial and lateral rods.  
We designed for 25\,pounds of axial load (equivalent to the
load on an axial rod)
and the full load
of the mirror laterally for each restraint.  
This system has a small safety factor; the restraints
may bend/deform
during an earthquake and would need 
to be replaced afterwards.

\subsection{Mirror Deployment (Swing arm)}
\label{sec:swing_arm}

\subsubsection{Design}

The K1DM3 system is designed to deploy and retract its mirror upon software command. In the following, we describe the parts critical to the actuation of K1DM3 with the exception of details on the bipods and 
kinematic couplings that position the assembly during deployment.
Those are discussed in $\S$~\ref{sec:upper}.

\begin{figure}[h!]
\begin{center}
\includegraphics[width=2.5in]{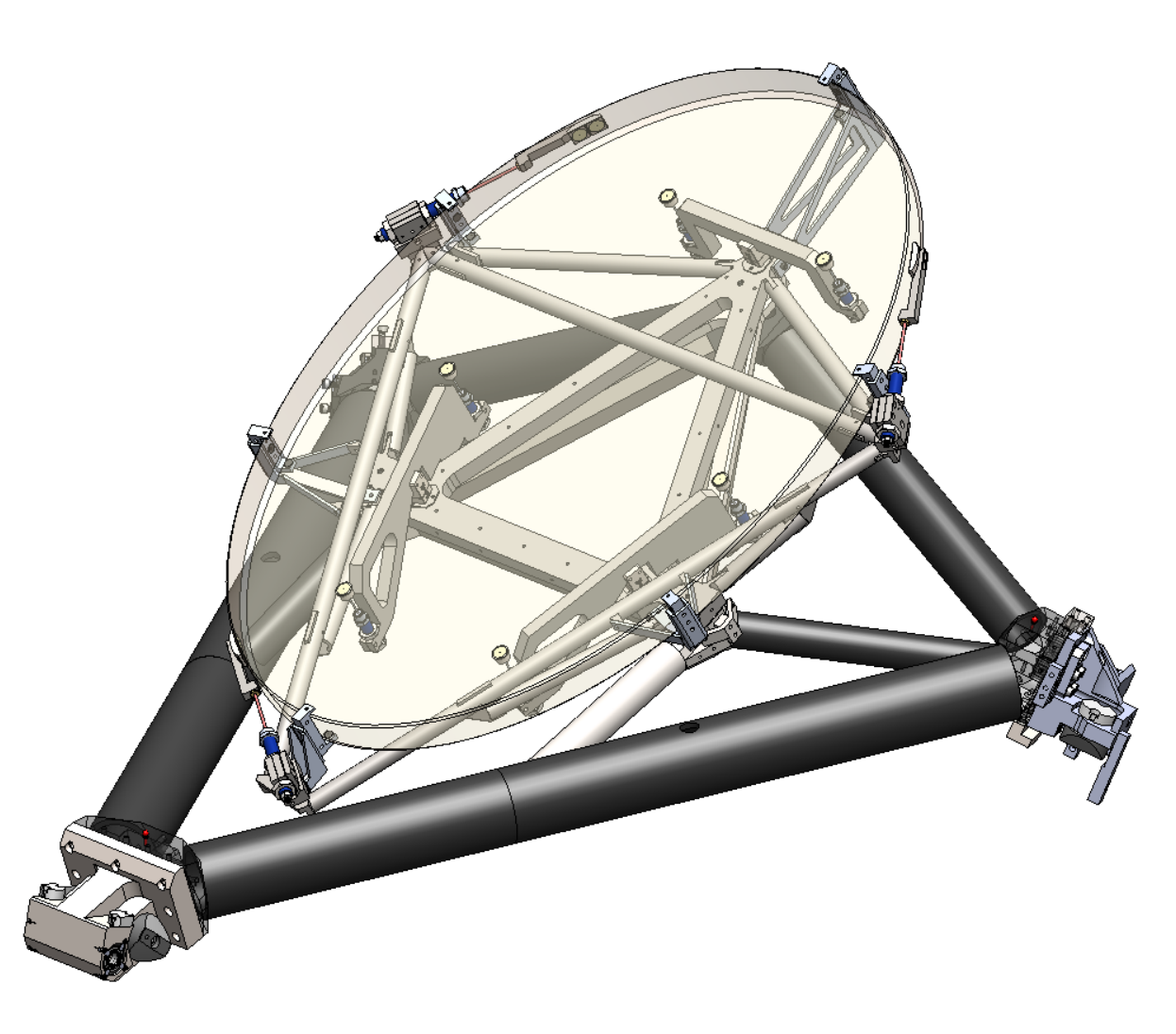}
\includegraphics[width=2.5in]{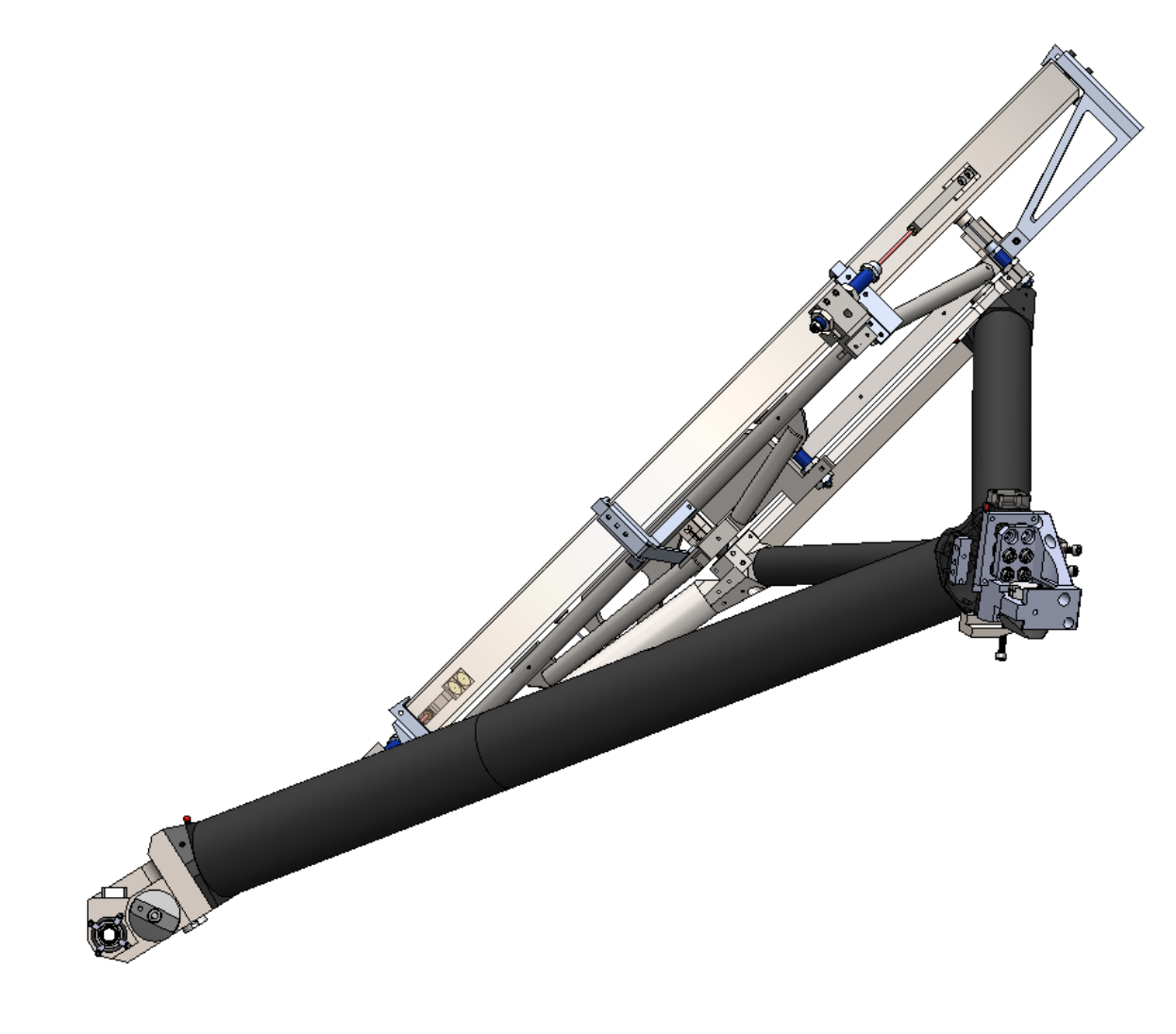}
 \caption[Swing arm]{\footnotesize 
Perspective and side views of the mirror support swing arm.
}\label{fig:swingarm}
\vskip -0.1in
\end{center}
\end{figure}

\begin{figure}[h!]
\begin{center}
 \includegraphics[width=3.0in]{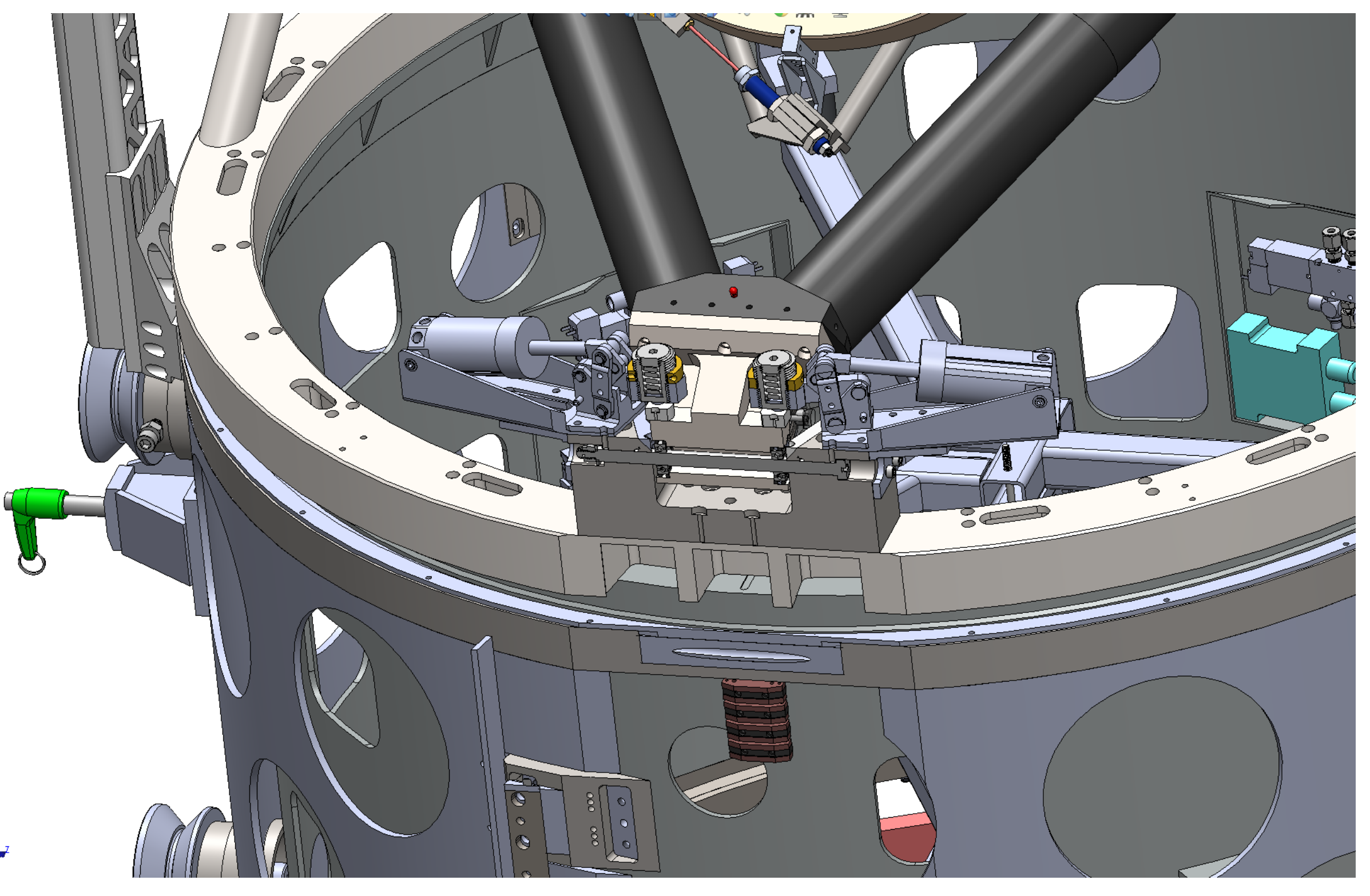}
 \caption[Pivot for the swing arm]{\footnotesize 
View of the swing arm pivot.
}\label{fig:pivot}
\vskip -0.1in
\end{center}
\end{figure}

The Mirror Assembly described in the previous section will fasten to a tripod swing arm fabricated with ASTM-A36 steel. Figure~\ref{fig:swingarm}
illustrates the shape and overall dimensions of this part. It is a weldment of several steel members with varying diameter designed to maintain
rigidity while minimizing profile. At the end points of the main arms are the kinematic couplings (canoe spheres made by Baltec, division of Micro Surface Engineering, Inc.) that enable repeatable, precise positioning of the system. This swing arm is attached to a pivot on the top ring of the K1DM3 module. This pivot is compliant ($\approx 1$\,mm)
to allow the kinematic coupling to determine the 
deployed position of the mirror. The pivot mechanism 
consists of a shoulder screw that is supported by 
two spherical-rolling element bearings, each supported by O-rings.
(see Figure~\ref{fig:pivot}).

The swing arm is pushed into place by a pair of linear actuators (Exlar model GSX40-0601). Each actuator is attached to the bearing ring by 
a pinned spherical joint. The opposite end is attached to the swing arm with a universal joint.

\subsubsection{Design Analysis:}

\noindent
\underline{Motor torque:}
The two linear actuators need sufficient torque to deploy and retract the Actuation assembly and Mirror assembly under gravity. 
The required torque 
will be less at the elevation angle designed for deployment
(\depang) and at the nominal
drum rotation angle of 90\,degrees (where 0\,degrees faces K1DM3
towards the AO system), but we have calculated the torque required assuming a worst-case configuration. Specifically, this implies a force of 6860 N. Each of the linear actuators has a manufacturer reported force of 9450 N.

\noindent
\underline{Bipod strut placement:}
When deployed, the K1DM3 mirror will be lowered onto three kinematic fixtures (grooves) held in a plane by weldment consisting of
two bipod struts on a large ring. 
These were positioned (i) to avoid vignetting the converging beam from M2 to the Cassegrain focus; (ii) to orient the three grooves in a plane (or parallel planes); (iii) to mount on the bearing ring; and (iv) to avoid collisions with the known extensions of the tertiary tower at
any rotation angle of the K1DM3 module
(e.g. Acme screw presenters, forward baffle mounts). 
The current design satisfies all of these constraints.

\noindent
\underline{FEA analysis:}
FEA models were made of the mirror assembly attached to the swing arm and connected to the bipods.  This led to an iterative process for the design of the
placement, size, and material of the swing arm struts.
The model accurately represents the mirror with its support by the axial and lateral flex rods. The swing arm structure was modeled 
as a series of beams with full elastic properties reflecting
the materials of construction.
The kinematic restraints at the v-groove are modeled as pin connections.
The bases of the bipods, which connect to the upper ring, 
are fixed to ground.  The pivot was also included.
Extensive analysis of the deformation of the mirror surface has been conducted by independent analysis covered earlier. The purpose of the model and analysis described here is to determine the performance of the supporting structure and rigid body displacement and rotation of the mirror.

We then measured the deflections in the mirror from nominal position
(telescope pointed at Zenith) for the extreme case
of observing at 72\,degrees off Zenith.
Results of the static gravity cases are shown in Table~\ref{tab:gravity}. Only the displacements causing out of plane motion 
of the mirror are reported. These are piston, tip (rotation about the minor axis), and tilt (rotation about major axis) of the mirror.
These values are satisfy the requirements for positioning
and repeatability of the mirror.

\begin{table}[h]
\begin{center}
\caption{Out of plane displacement and rotations to due gravity for the
mirror when observing at 72\,degrees off Zenith.}
\label{tab:gravity}
\begin{tabular}{cccc}
\hline
{\bf Focus} & {\bf Piston ($\mu$m)} & {\bf Minor axis ($''$)} &
{\bf Major Axis ($''$)}\\
\hline
bent-Cass & 114 & 2.4 & 3.6 \\
AO        &  64 & 9.1 & 5.7 \\
\hline
\end{tabular}
\end{center}
\end{table}

\noindent
\underline{Vibrational analysis:}
Over the course of its lifetime, the K1DM3 module will experience a
range of stresses as it is transported to WMKO, handled during
installation and coating, and tipped about during observations.
As such, it must be designed to survive these conditions and maintain
sufficient stability during observations to not significantly
degrade the image quality.

There are two particular requirements that the K1DM3 design must
satisfy regarding vibrations and the stresses that result:
(1) The stresses on the mirror must not compromise the structural
  integrity of the K1DM3 components.  Our primary concern is the glass
  of the mirror which has a maximum tensile strength for Zerodur of
  42~MPa (6000~PSI) with a vendor-recommended limit of 10~MPa. 
  We have designed K1DM3 to limit stresses to be less than 1000~PSI.
(2) Motions of the mirror must contribute less than 10\%\ rms to the
  optimal seeing disk of 0.4\,arcseconds.  As described in 
  \cite{k1dm3_positioning}, 
  this implies less than 29 microns (RMS) of
  translational motion and less than 0.65\,arcseconds and 0.46\,arcseconds (rms) for
  motions in tip and tilt respectively.

We examined the predicted
performance of the current K1DM3 design in a range of environments.
Specifically,
the K1DM3 team was provided a set of forcing functions 
(Tables 4,5,6 of \cite{k1dm3_requirements})
that are
intended to describe the random motions as a function of frequency
that the module would experience in three conditions: 
 (1) during transportation;
 (2) during handling at WMKO (e.g.\ installation, removal for coating);
 (3) during normal operations when installed on the telescope.

The maximum stresses for the K1DM3 mirror when deployed are
444\,PSI, 163\,PSI, and 0.1\,PSI for transportation, handling,
and observing respectively.  
Not surprisingly, the stresses on the mirror during transportation
would nearly exceed our 500\,PSI limit.
We will design the shipping crate and packing materials 
to reduce the stress to tolerable levels.
Regarding normal operations when deployed, we estimate 
$0.3$\,arcseconds (rms) of tilt and tip motions and 1\,micron of
translation.  The stress of 0.1\,PSI is far within tolerance
and we predict maximum accelerations of less than 0.05\,g.

\noindent
\underline{Vignetting:}
$\S$~\ref{sec:vignette} 
provides a discussion of vignetting of the beam by K1DM3. 
Figure~\ref{fig:retract} shows that neither the struts nor the retracted mirror vignettes the LRIS or MOSFIRE footprints.
We are required, however, to rotate the mirror to face one corner
of the hexagonal face of the tertiary tower to avoid the beam
from M2 to M1 (when the secondary baffle is not installed).
This is the nominal parking position for K1DM3 when retracted.

\noindent
\underline{CTE:}
The temperature of the entire telescope will change by as much as 25\,degrees C (summit temperatures vary from 14\,deg C to $-11$\,degrees C). Given this temperature variation, it is important to consider thermal expansion effects on the alignment of the tertiary mirror.

Our goal has been to limit the effects from thermal expansion to be the same or less than those experienced by the current M3 system. The key to reducing the sensitivity is to pick materials with coefficients of thermal expansion that match the rest of the telescope structure. The predominant material used in the telescope structure is ASTM-A36 steel which has a CTE of 11.7 ppm/\,degrees C. This material and material with very similar CTEs have been selected for the K1DM3 design 
We estimate that there will be an approximately
.12 mm change in the height of 
the struts for 25\,degrees C, but all three will move together 
maintaining the geometry.

\noindent
\underline{Time to Deploy/Retract:} 
With the dual actuator design, the rod must retract approximately 100\,mm 
to change from the retracted position to deployed. The manufacturer-listed peak speed is 21\,mm per second. We will actuate slower than the peak speed and allow 30 seconds to provide a smooth velocity profile for both deploy and retract.

\begin{figure}[h!]
\begin{center}
 \includegraphics[width=2.5in]{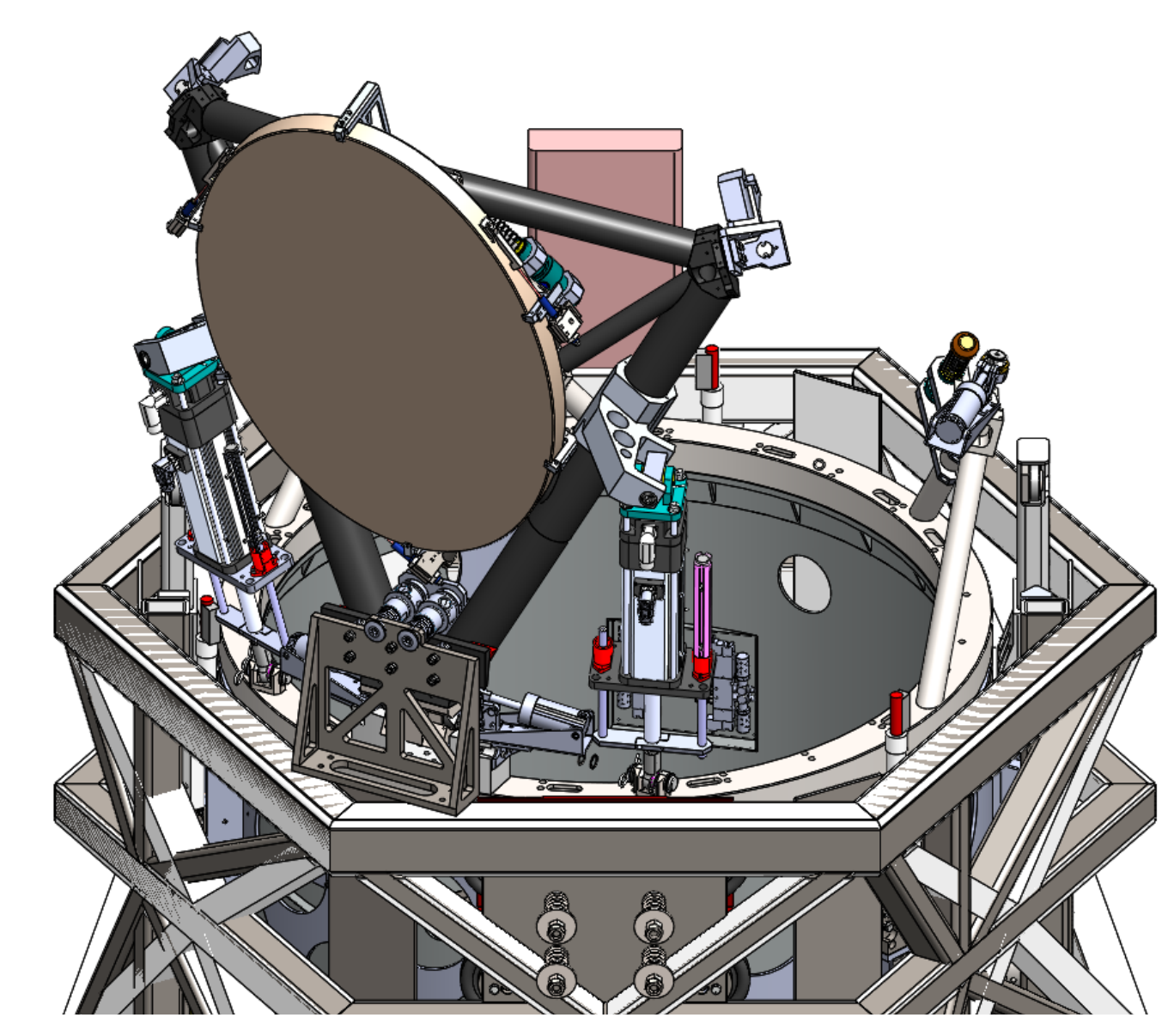}
 \includegraphics[width=2.5in]{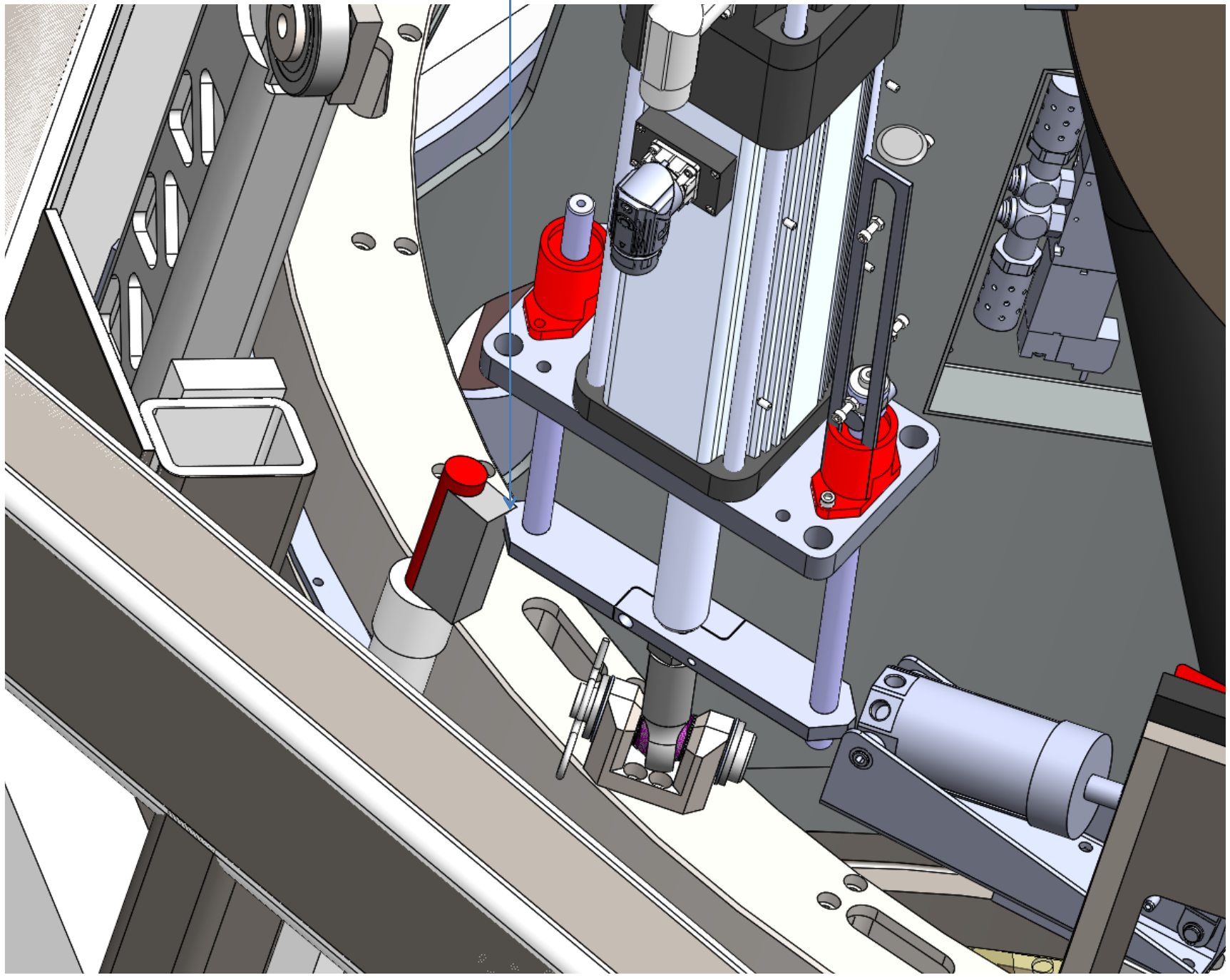}
 \caption[Parking position for K1DM3 retracted]{\footnotesize 
(left) Parking position of the K1DM3 mirror when retracted.  In
this position, no portion of the assemblies vignettes the
light traveling to the Cassegrain instruments nor the light
from M1 to M2.  
(right) Zoom in on potential interference between K1DM3
and the magnetic switch on one of the Acme screw presenters
of the DMP system.  We may need to rotate these switches to
avoid interference.
}\label{fig:top_view}
\end{center}
\end{figure}

\noindent
\underline{Tower clearance:}
To avoid vignetting in certain configurations of the M2 baffles, 
we must rotate the K1DM3 mirror when
retracted to the parking position indicated in Figure~\ref{fig:top_view}.
During Detailed Design, WMKO approved and executed the removal of
the forward baffle tracks on the K1 tertiary tower.  This reduced
the potential for interference between K1DM3 and the tower.
The only remaining source of significant, potential interference
are the Acme screw presenters 
associated with the tower defining points.
We have modeled in SolidWorks the top of the tertiary tower using the WMKO drawings
and measurements by our team during DD. 
We have also modeled
the three mechanisms for the telescope half of the tertiary module/K1DM3 defining points that are located 659.5\,mm 
from the optical axis, 35 mm diameter in diameter, 
and extend 100 mm above the tower.  

We then confirmed using SolidWorks that the swing arm assembly 
of the K1DM3 design clears all of these obstructions when rotating to the parking
position.
This required us to reduce the profile of 
the Exlar actuators and use a particular orientation for mounting
these devices.
There is a potential interference with the magnetic switches
on the cylinders (Figure~\ref{fig:top_view}, right) which may
require us to rotate these prior to installation of K1DM3.


\subsection{Upper Assembly Design}
\label{sec:upper}

\subsubsection{Design Description}

\noindent
\underline{Upper ring:}
At the top of the two drums is an upper ring 
(structural steel) which
supports the swing arm (described in 
$\S$~\ref{sec:rotation}).  This part has an inner
diameter of 1104\,mm, an outer diameter of 1240\,mm,
and is 56.4\,mm wide. 
It is manufactured with a recess that centers it
on the inner drum.

\noindent
\underline{Bipod struts:}
Also attached to the bearing ring is a pair of bipod struts made of A500 tubular steel.  Each bipod holds a v-groove kinematic coupling 
for positioning the mirror when deployed. The struts are approximately 457\,mm long and are positioned at 120\,degrees from the swing arm pivot. These are tilted at an 
angle 13.7\,degrees away from the optical axis. 
The third sphere/v-groove interface is adjacent to the compliant point.

\begin{figure}[h!]
\begin{center}
 \includegraphics[width=2.5in]{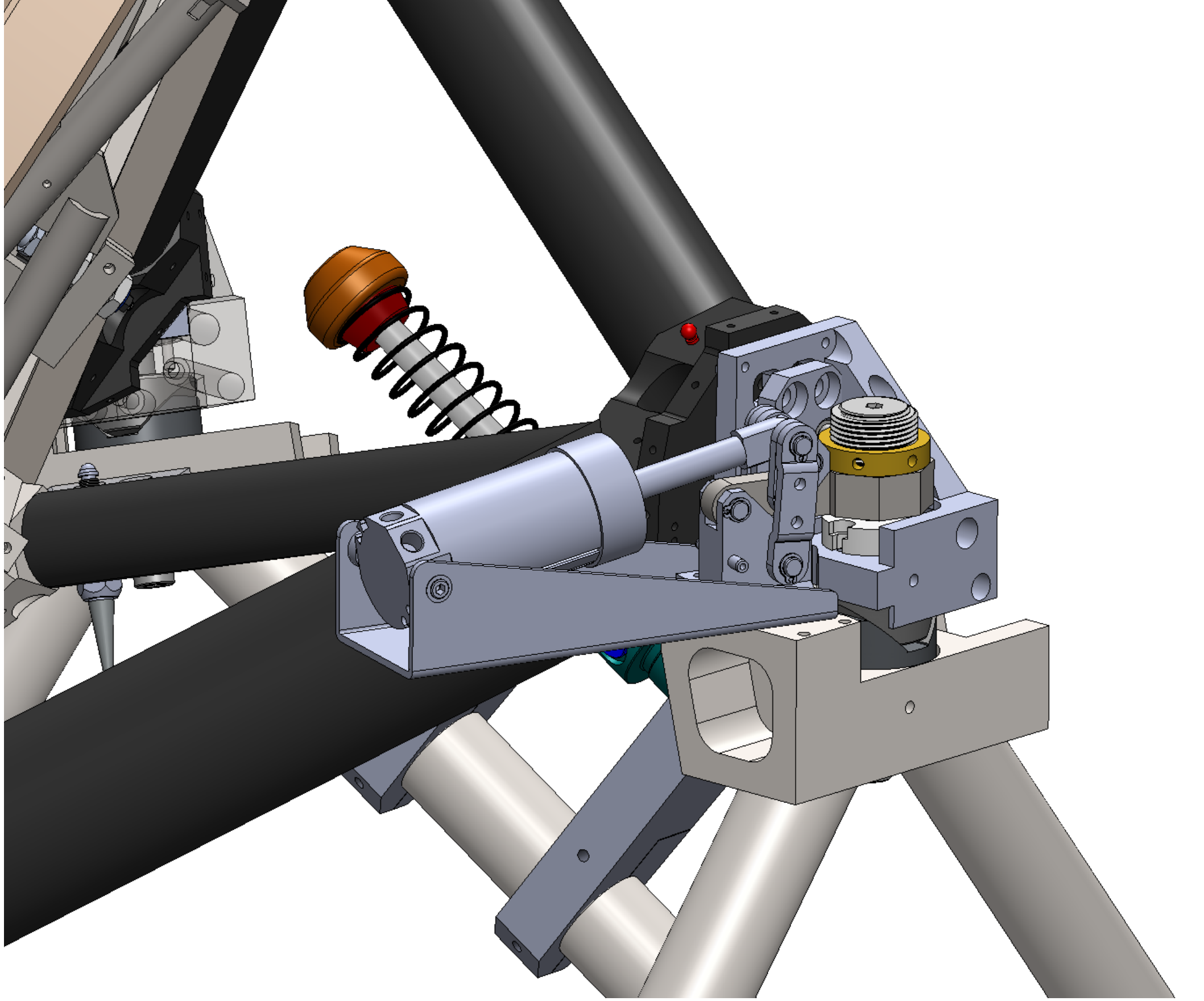}
 \caption[Dampers near deploy kinematics]{\footnotesize 
 In the foreground, one views the damper attached to the
 bipod strut which will slow the K1DM3 swing arm in the event
 of a failure of the actuators.  The clamp mechanism which
 holds tight the kinematic coupling at the top of the bipod
 struts is also shown.
}\label{fig:damper}
\vskip -0.1in
\end{center}
\end{figure}

As a safety measure, we will attach an Enidine damper to each of the bipod 
struts to ``catch'' the swing arm assembly in the event of a major 
software or hardware failure.  This will insure that the kinematics 
engage with the swing arm moving at a speed of less 
than 10\,mm per second.  
Figure~\ref{fig:damper} shows the designed damper system.

\noindent
\underline{Deploy kinematics (DKs):}
We will employ 3 sets of canoe-sphere/v-groove fixtures 
in the kinematic coupling used for positioning the K1DM3 mirror when deployed. The canoe spheres will have a radius of 500\,mm. Two of these are mounted at the ends of the swing arm and the third is mounted on the under-side of the swing arm at approximately 44\,mm from the pivot point.

The v-grooves will be 25.4 mm wide and 49 mm long. Two of these will be held at the ends of the bipod struts in a common plane. The third v-groove is mounted to the end of the pivot mechanism for the swing arm. Its axis lies in a plane parallel to that defined by the struts but offset by 655\,mm.
All of the DK fixtures will be made with 440 stainless steel, 
polished to 0.2 micrometer rms roughness, and plated 
with chromium nitride and tungsten di-sulfide (WS$_2$).
The former is to prevent rust and the latter 
is to achieve a low coefficient of friction.

\noindent
\underline{Clamps:}
When engaged the DKs will be clamped with a pneumatic
clamping mechanism that maintains 1500\,N of sustained forced 
on each canoe sphere/v-block interface 
(Figure~\ref{fig:damper}). 
There is 1 clamp on each bipod strut and a pair at the pivot
(Figure~\ref{fig:pivot}).
These air actuated mechanisms will be controlled via 
a solenoid valve which holds the coupling 
in place even with a loss of air
pressure. It will be important for the actuation assembly to bring the kinematic fixtures in 
close contact, but this pneumatic mechanism will be 
relied upon to fully engage the coupling.
Feedback switches will be provided to verify that 
each clamp is open or closed.

\subsubsection{Design Analysis:}

\noindent
\underline{Deploy Kinematics (DKs):}
After studying the standard reference on kinematic couplings \cite{slocum92}, we decided on a canoe-sphere/v-groove coupling system for the DKs. These have advantages over the more traditional cone-flat-groove couplings. We then researched vendors that manufactured these fixtures and selected Baltec. We have analytically estimated the precision for repeatable positioning as described in 
\cite{k1dm3_positioning}. 

Because the DKs are critical to the success of K1DM3, we purchased a complete set of fixtures during PD and manufacture test beds to construct a kinematic coupling. We then tested the positional performance empirically with three LVDTs. 
We will perform a series of new positioning and repeatability 
tests with the new set of kinematics at the start of full
scale development.

\noindent
\underline{Clamping:}
For positional stability, each DK will be clamped with a sustained force of 1500\,N by a clamping mechanism (Figure~\ref{fig:damper}). 
We estimate less than 1.0 microns of motion with any change of gravity. This satisfies the requirements on stability.

\noindent
\underline{Damper system:}
We selected a set of Enidine damper devices (OEMXT 1.5Mx3)
that have a damping
force that will slow the K1DM3 swing arm assembly to less
than 10\,mm per second
in the event of a complete loss of the actuators.
This speed was estimated assuming the total weight of 
the swing arm assembly is borne by the dampers.
These were also chosen because of their small profile,
i.e.\ to avoid vignetting the light path.

\subsubsection{Prototyping:}

We have the as-built kinematic couplings at UCO and 
will test their performance with an upgraded swing
arm during FSD.
We have built one of the clamping mechanisms at UCO
and have tested its performance.  Provided it is driven
with air pressure exceeding 90\,PSI, we achieve 
1500\,N of force.

\subsubsection{Fabrication:}

The deployable kinematics were made by Baltec
and then polished at UCO.  These were then coated with 
CrN and WS$_2$.
The clamping mechanisms have been purchased
from DeStaCo and modified at UCO.  We have two complete
systems in operation.
The bipod struts will be fabricated
an external vendor.

\begin{figure}[h!]
\begin{center}
 \includegraphics[width=4.0in, angle=-90]{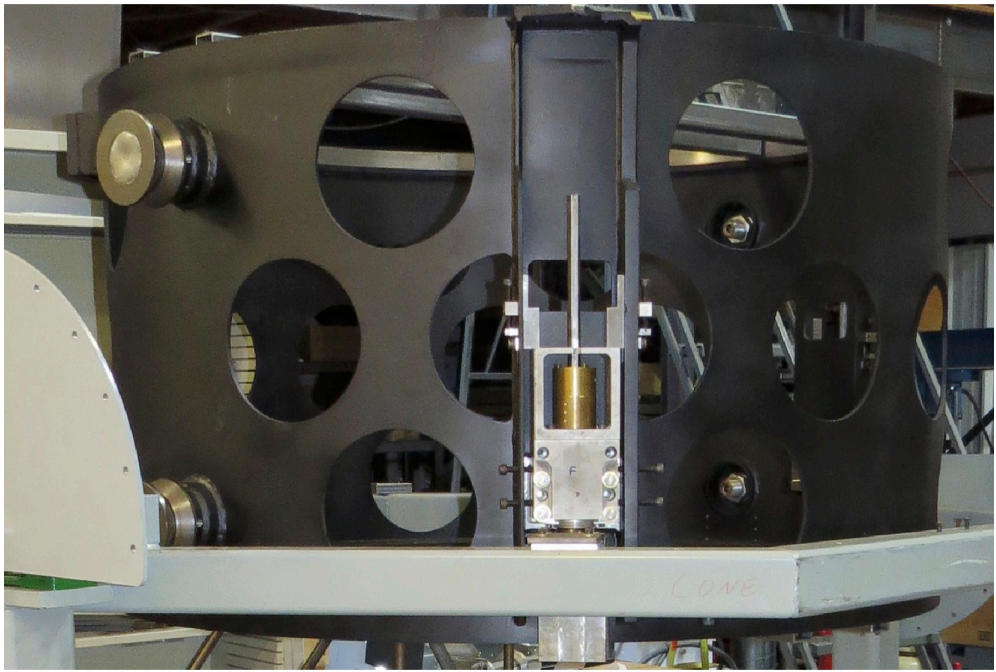}
 \caption[Outer drum]{\footnotesize 
Photo of the fabricated outer drum for K1DM3 (at UCO).
}\label{fig:outer_drum}
\vskip -0.1in
\end{center}
\end{figure}

\subsection{Drum Assembly Design}
\label{sec:rotation}

\subsubsection{Design Description:}

\noindent
\underline{Outer drum:}
The backbone of the K1DM3 system is an ASTM A36 steel drum, 1/4\,inch
rolled plated precision machined to have an outer diameter of 1240\,mm. 
The drum has holes to reduce its mass without compromising its stiffness 
(Figure~\ref{fig:outer_drum}). 
The drum is approximately 737\,mm long.

\noindent
\underline{Wheels:}
For installation and removal of K1DM3 in the K1 tertiary tower,
we have purchased (from Osborn) 4 steel wheels 
each with a diameter of 4.5\,inch.
These are now attached to the outer drum (Figure~\ref{fig:outer_drum}).

\noindent
\underline{Anti-tip arms:}
As designed, the center of mass of the K1DM3 module when
the mirror is deployed lies 63\,mm behind the forward wheels
(those closest to the mirror) toward the rear set of wheels.
In principle, this implies the system is stable to tipping.
To further mitigate against tipping during installation, 
we have designed bronze counter-weights attached to the
lower end of the assembly
and a pair of anti-tip arms which can bear the full weight of the system.
These will be affixed to the outer drum as an
additional safety measure.

\noindent
\underline{Defining points:}
Fixtures attached to the exterior of the K1DM3 module to position it within the tertiary tower. These are referred to as the Defining Point Mechanism (DPM), and the complete kinematic coupling includes the tower fixtures. These provide the mechanical interface between K1DM3 and the telescope.
The DPM fixtures on K1DM3 replicate the fit and function of the fixtures on the current tertiary module. We re-used the existing design
to the maximum extent possible. There are three fixtures located with 120\,degree separations around the module, approximately 
141\,mm 
below the top surface of the top bearing with three different contact points to form a kinematic mount: (i) Flat on flat; (ii) Sphere in cone; and (iii) Cylinder in groove.

Each defining point mechanism consists of two parts or halves. One half of each defining point mechanism is mounted on the tertiary tower and incorporates a rotationally fixed Acme thread lead screw that is extended through the fixed half of the kinematic point by an air cylinder when the defining sequence is initiated. This ``presents'' the lead screw to the instrument mounted half of the defining point which has a hole in the center of the mating half of the kinematic point. Behind this is an Acme thread nut which engages the fixed lead screw presented by the tertiary tower half of the defining point mechanism. The nut is rotated by a reversible air motor incorporated in the instrument half of the defining point mechanism. Once the two halves of the three defining points are all in initial contact, each defining point is mated in sequence, starting with the sphere, then the cylinder, and then the flat. The system is very tolerant of small misalignments, and each defining point can carry loads in excess of 2000 kg. Position sensors are incorporated in the system to ensure proper positioning before the defining sequence is started.

Figure~\ref{fig:outer_drum} shows
the as-built sphere-in-cone DPMs mounted on the outer drum of K1DM3.
Each of the DPMs are made of AISI 4130, Hardness Rockwell C 50 minimum, chrome plated to QQ-C-320, Type 1, class 2 material. Each allows for adjustment for 6 degrees of freedom (3 translational and 3 angles). These will have an end-to-end positioning range of about 12\,mm.

\noindent
\underline{Ring bearings:}
Attached to the top and bottom of the drum are two ring bearings with a nominal ID of 1180\,mm and 1218\,mm OD. They are essentially identical in form, fit, and function to the bearings on the existing modules.  These enable the inner drum assembly to rotate about the optical axis. The top bearing will be located approximately 711\,mm below the elevation axis. These were manufactured by Kaydon and have been delivered to UCO.

\noindent
\underline{Inner drum:}
The inner drum will be rolled 1/4\,inch plate, welded and turned ASTM A36 steel.
It will have an inner diameter of 1150\,mm, 
and be 768\,mm long.
The bearing surfaces will be concentric to 50\,microns
and the drum has an interference fit with the
supporting bearings of 50\,microns.  This interference
fit is to insure our rotation axis does not shift with
respect to the bearings.
We have also designed pockets for the Galil controller and
a programmable logic controller (RIO) which 
acts as an interlock device for the Galil.

\begin{figure}[h!]
\begin{center}
\includegraphics[width=2in]{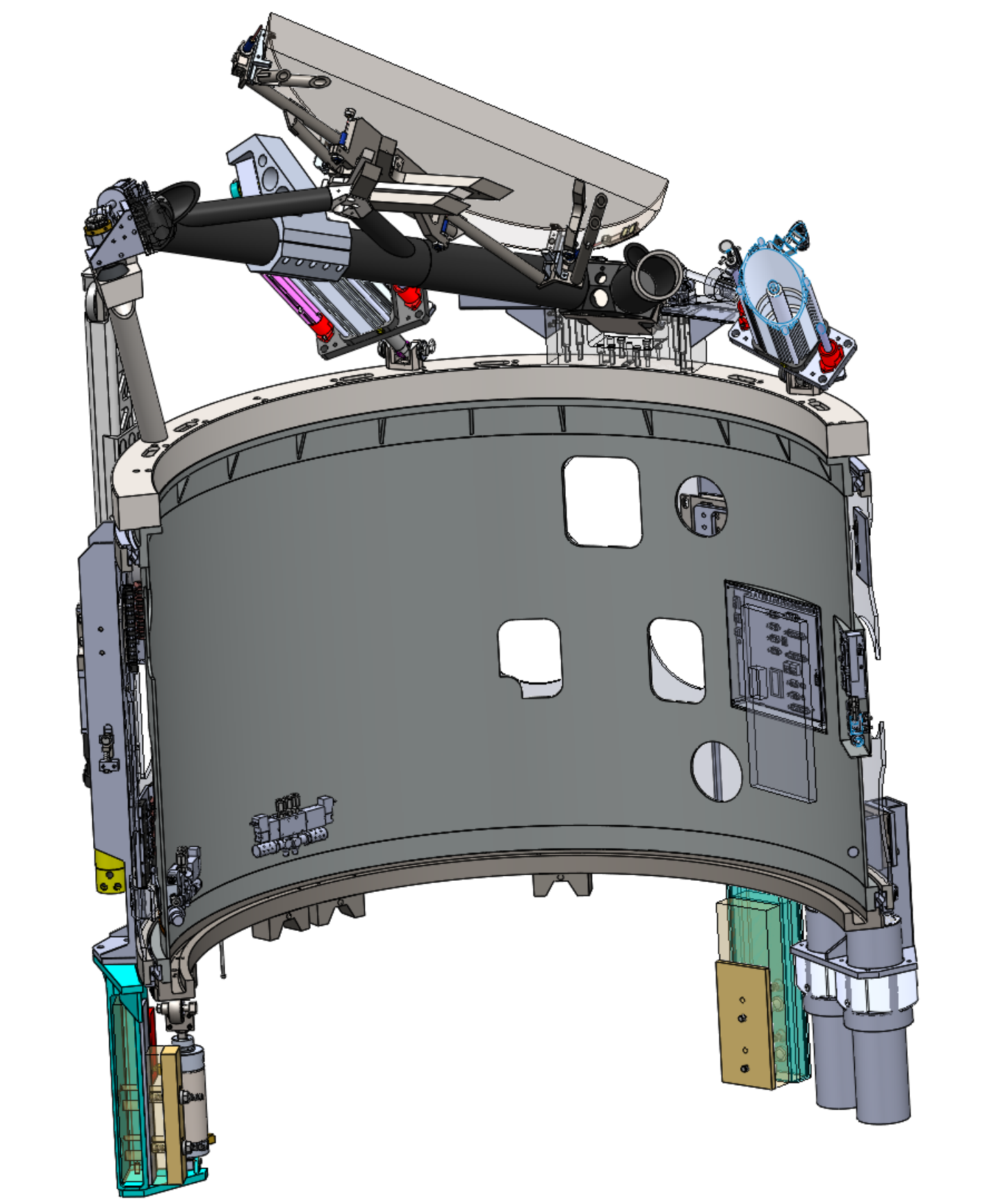}
 \caption[Cross-section of inner drum]{\footnotesize 
Cross-sectional view of the inner drum.  One notes
``pockets'' for the Galil and RIO mechanisms and fixtures
for air supply and electric connectivity.
}\label{fig:inner_drum}
\vskip -0.1in
\end{center}
\end{figure}

\begin{figure}[h!]
\begin{center}
\includegraphics[width=2.0in]{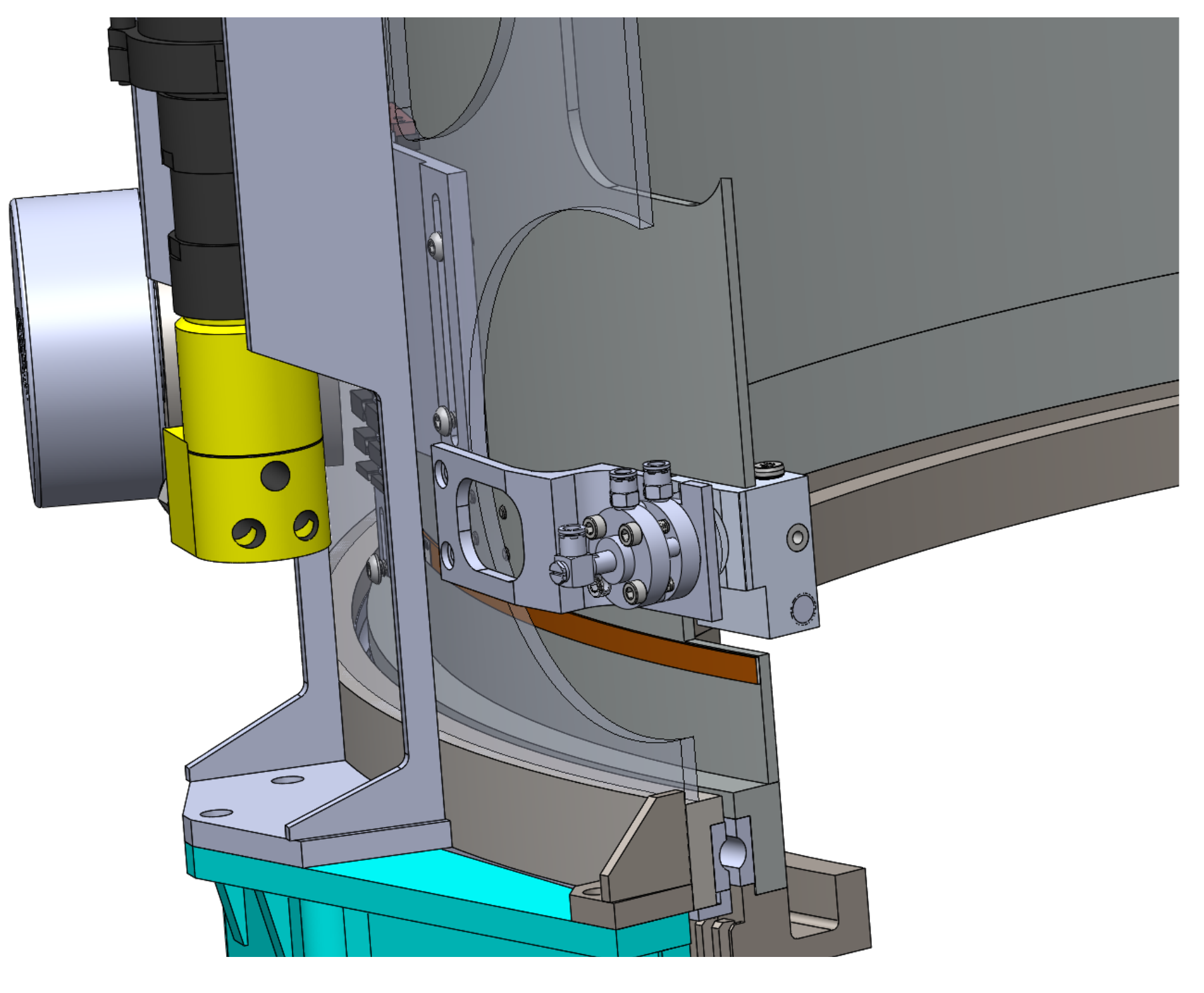}
\includegraphics[width=2.0in]{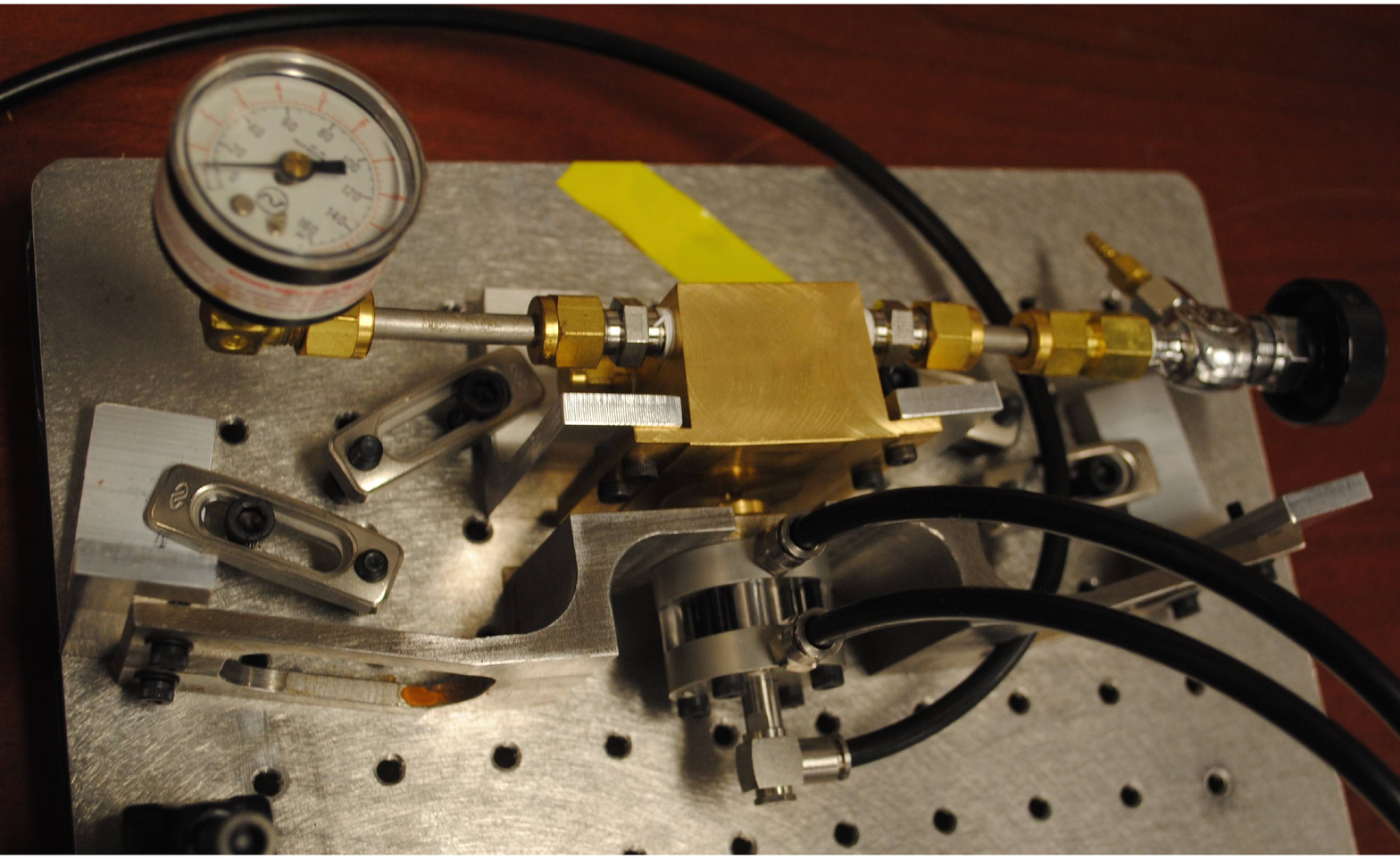}
 \caption[Compressed air fixture]{\footnotesize 
 Figure showing the fixture supplying compressed air to the
 upper assembly (top) and a photo
 of the prototype of the fixture (bottom).
}\label{fig:pneumatics}
\vskip -0.1in
\end{center}
\end{figure}

\noindent
\underline{Compressed Air supply:}
To provide air pressure to the pneumatic clamps, we have designed
a custom fixture that connects an air supply on the fixed, outer
drum to plumbing on the inner drum. 
There will be a pair of these mechanisms at 0 and 180\,degrees rotation
angles.  The latter is the orientation when the
mirror is deployed/retracted and the former is for 
removing the mirror for coating.
Figure~\ref{fig:pneumatics} shows one device in the
assembly and a photo of our prototype.
The system will engage when oriented to within 1\,degree
of alignment.

\noindent
\underline{Ring gear and detents:}

\begin{figure}[h!]
\begin{center}
\includegraphics[width=2in]{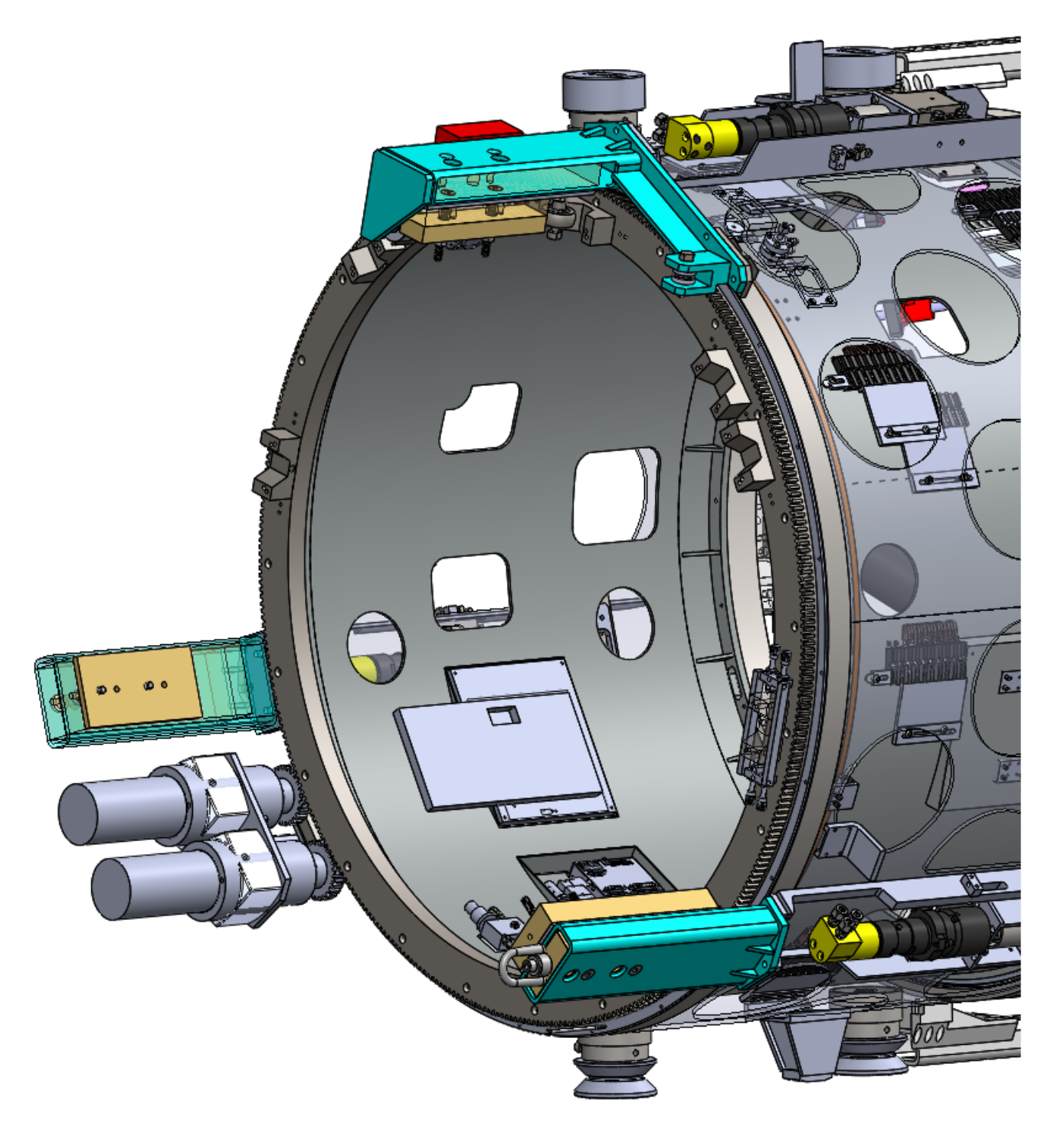}
 \includegraphics[width=2in]{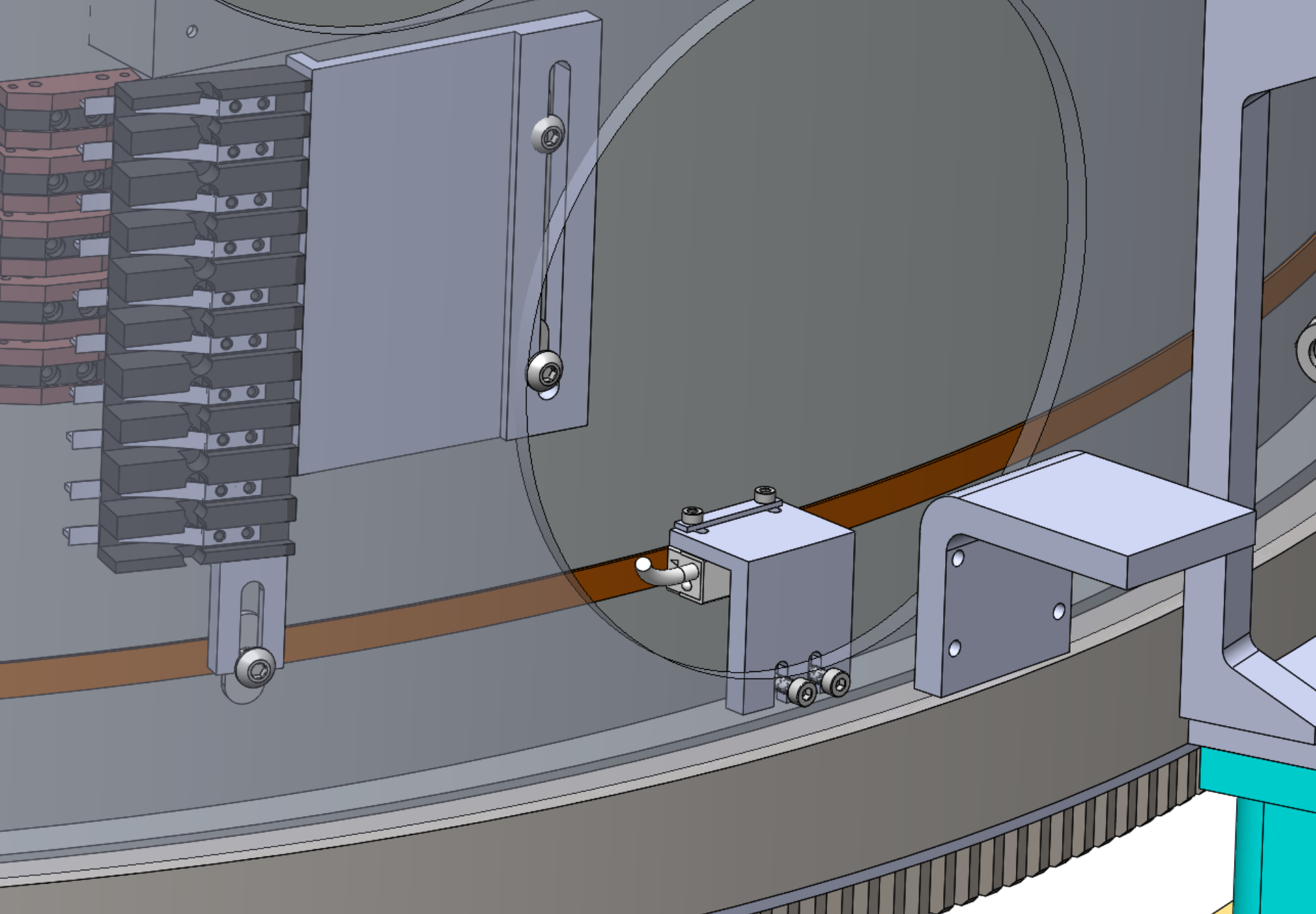}
 \caption[Ring gear]{\footnotesize 
(left) Ring gear, detents (v-grooves) and detent mechanism (cyan, top).
(right) Renishaw tape (brown) on the inner drum.
The figure also shows the readhead in the assembly.
}\label{fig:ring_gear}
\vskip -0.1in
\end{center}
\end{figure}

As shown in Figure~\ref{fig:ring_gear},  the very bottom will be a ring gear and servo system to position the inner drum to within 10 microns. Six detents (v-grooves) are mounted to this ring to precisely set the rotation angle for the foci on the elevation ring. 
The ring gear will be made of ANSI 4130 steel. 
There are two additional detents for deployment and retraction.
4130 material, have an inner diameter of 
1075\,mm, and 378 teeth around the 
full circumference. Two pinions gears turned by two DC servo motors using harmonic drive gear heads will drive the ring gear. One of the two servo motors will lag the other slightly to eliminate backlash between the drive pinions and the ring gear.

There will be eight detent blocks
made of steel hardened to 45 to 50 Rockwell Scale C and will be v-grooves that are 88\,mm long and 40\,mm wide. These will be pinned to the ring gear during alignment at WMKO.
This coupling will be engaged by an air-pressure driven detent mechanism mounted below the module (see Figure~\ref{fig:ring_gear}).

\noindent
\underline{Encoder:}
The rotation angle of the K1DM3 module will be monitored by reading a 
Renishaw 
magnetic tape attached to the drum, approximately 78\,mm 
above the bottom edge as noted in Figure~\ref{fig:ring_gear}. 
We will use a single read head with an incremental encoder 
with distance coded reference marks
for precise and continuous reads.

We estimate a total mass for the K1DM3 drum and 
attached components of $\sim$595\,kg.

\subsubsection{Design Analysis:}

\noindent
\underline{Defining Point Mechanisms (DPMs):} 
By replicating the existing design for the K1DM3 defining point mechanisms, it is our expectation that the system will position as precisely as the current module. We estimate the positioning repeatability to be within 3 microns (lateral) and that the normal of the mirror will rotate less than 0.5\,arcseconds. 

During Detailed Design, we fabricated these fixtures and the outer drum. We delivered them to WMKO to test clearance and perform initial alignment of these fixtures (e.g. test that they coupled to the tower fixtures).
We successfully mounted the outer drum on the K1 tertiary tower defining
points and measured that its center of rotation lay within several mm
of the current tertiary module (see \cite{k1dm3_ODF}
for further details).

Our design will allow for 12 mm of positioning of the DPMs
to facilitate the alignment of K1DM3.

\noindent
\underline{Stiffness:}
As the underlying support for the K1DM3 system, the drum must be sufficiently stiff and strong to hold the Actuation Assembly in place under a varying gravity vector. We have estimated the flexure in the drum by performing an FEA within SolidWorks. We estimate that the deflection between vertical and horizontal orientations is 0.65\,arcseconds and approximately 
1.1 microns of translation. These are primarily due to flexure of the two bearings. Such deformations are within the requirements for positioning.

\noindent
\underline{Vignetting:}
Unlike the drum of the current tertiary module, the Drum assembly of K1DM3 must be sized to avoid vignetting the converging beam from M2 to the Cassegrain focus. For example, the ring gear designed for the current existing tertiary module has too small of an inner diameter and we have re-designed it accordingly. $\S$~\ref{sec:vignette} describes our vignetting analysis in detail and we find that the Drum assembly does not vignette the beam at any rotation angle.

\noindent
\underline{CTE:}
The drum will have the same CTE as the
tower and existing tertiary module.
This mitigates against the effects of thermal expansion.

\noindent
\underline{Anti-tip:}
The center of mass of K1DM3 with the mirror
deployed is estimated to lie 63\,mm behind the forward set
of wheels toward the back set of wheels.
Each anti-tip arm was designed to hold the entire force
of the K1DM3 module (3000\,N)
if it tipped and expressed the entire force
onto only one rail (allowing for misalignment
of the rails).  We estimate a maximum stress of 12510\,PSI
which is a factor of 4
below the yield stress of each anti-tip arm.

\noindent
\underline{Rotation analysis:}
We may estimate the rotation speed of the K1DM3 module as follows. The peak capable speed will be 18\,degrees/s if we run the servo at 1079 RPM, given the gear reduction of 354:1. We will limit this speed to a lower number (e.g. 9\,degrees/s) which would still allow K1DM3 to move from AO to HIRES in 20 seconds. The velocity profile will be a trapezoidal shape and utilize features such as S transitions to minimize vibration. The K1DM3 module may be rotated to any angle 
with the mirror deployed.  Interference with components on the
tertiary tower limits the rotation to approximately
30\,degrees when retracted. 

\noindent
\underline{Rotation positioning:}
The previous section described the positioning of the K1DM3 module in the tertiary tower using the Module Kinematics. Regarding rotational position of the Drum, the K1DM3 module will have eight detents (v-grooves) bolted to the bottom ring bearing. These are to be positioned such that K1DM3 precisely folds the light from M2 into the Nasmyth and bent-Cass foci. The precise location of the v-grooves will be established during Alignment at WMKO. The detent mechanism provides sufficient force (1200 N) to 
insure the module is locked into place and also provides stable
orientation for mirror removal and the mirror retract and parking positions. 
This pneumatic mechanism also reliably releases the detent.

\subsubsection{Prototyping:}

We had the outer drum fabricated during DD
and delivered it to WMKO to test its integration \cite{k1dm3_ODF}.
This includes our coupling the K1DM3 defining points
to the K1 tertiary tower defining points.
For the DPMs, we have mounted the outer drum with the
fabricated K1DM3 DPMs on the K1 tertiary tower in October 2015.
We have fabricated at UCO the custom fixture for providing
air to the clamping mechanisms.  This has been tested
under the environmental conditions (e.g.\ low temperature)
relevant to WMKO.

\subsubsection{Fabrication:}

The outer drum was manufactured by Wilcox.  
The wheels were made by Osborne and the
anti-tip mechanism was manufactured by UCO.
The pair of rotation bearings were manufactured by Kaydon.
The defining point mechanisms were manufactured
at UCO and have been attached to the outer drum.
The inner drum is being manufactured by Wilcox.
We have manufactured the pneumatic
coupling fixtures at UCO.  One unit is complete.
The ring gear will be fabricated by an external
vendor.  We have received quotes from Cage Gear and Machine
and Wilcox.
The detent mechanism and detents
will be manufactured by UCO.


\appendix    
\section{Electronics and Software} \label{sec:elec}


This Appendix provides a brief summary of the detailed design
of K1DM3 as regards electronics and software.

\subsection{Requirements}
\label{sec:electric_req}

\begin{my_enumerate}
\item  The K1DM3 system shall be powered from 120 Vac, 60 Hz. power at a maximum of 15 A. 
\item  The K1DM3 shall provide an emergency stop input that stops all motion when the Observatory emergency stop signal is activated. 
\item  The K1DM3 module shall not produce stray light from LED or lamp indicators, optical switches or optical shaft encoders over the wavelength range of 300 to 20000\,nm. 
\item  Cables and wiring shall be routed so that they do not interfere with the optical path of the telescope. Cables and wiring shall be routed so that full travel of moving or adjustable parts is not affected and does not place a strain on the mounting or connections of any cables or wiring. 
\item  The K1DM3 software user interface software shall be implemented as a DCS control row or other Observatory user interface paradigm. The user interface shall control the K1DM3 via keywords. 
\item  The K1DM3 software shall be written to run under a WMKO approved operating system. 
\item  The K1DM3 software shall be implemented as client-server architecture with communications over TCP/IP. 
\item  The K1DM3 software shall support legacy (current Keck telescope DCS) and new (TCSU) use cases. 
\end{my_enumerate}

\subsection{Electronics Design}

The electronics for K1DM3 provides control and feedback for four actions: rotating the drum, locking the drum position, deploying and retracting the mirror, and locking the mirror kinematics.

\subsubsection{Deployment Stage}

\begin{figure}[h!]
\begin{center}
\includegraphics[width=3in]{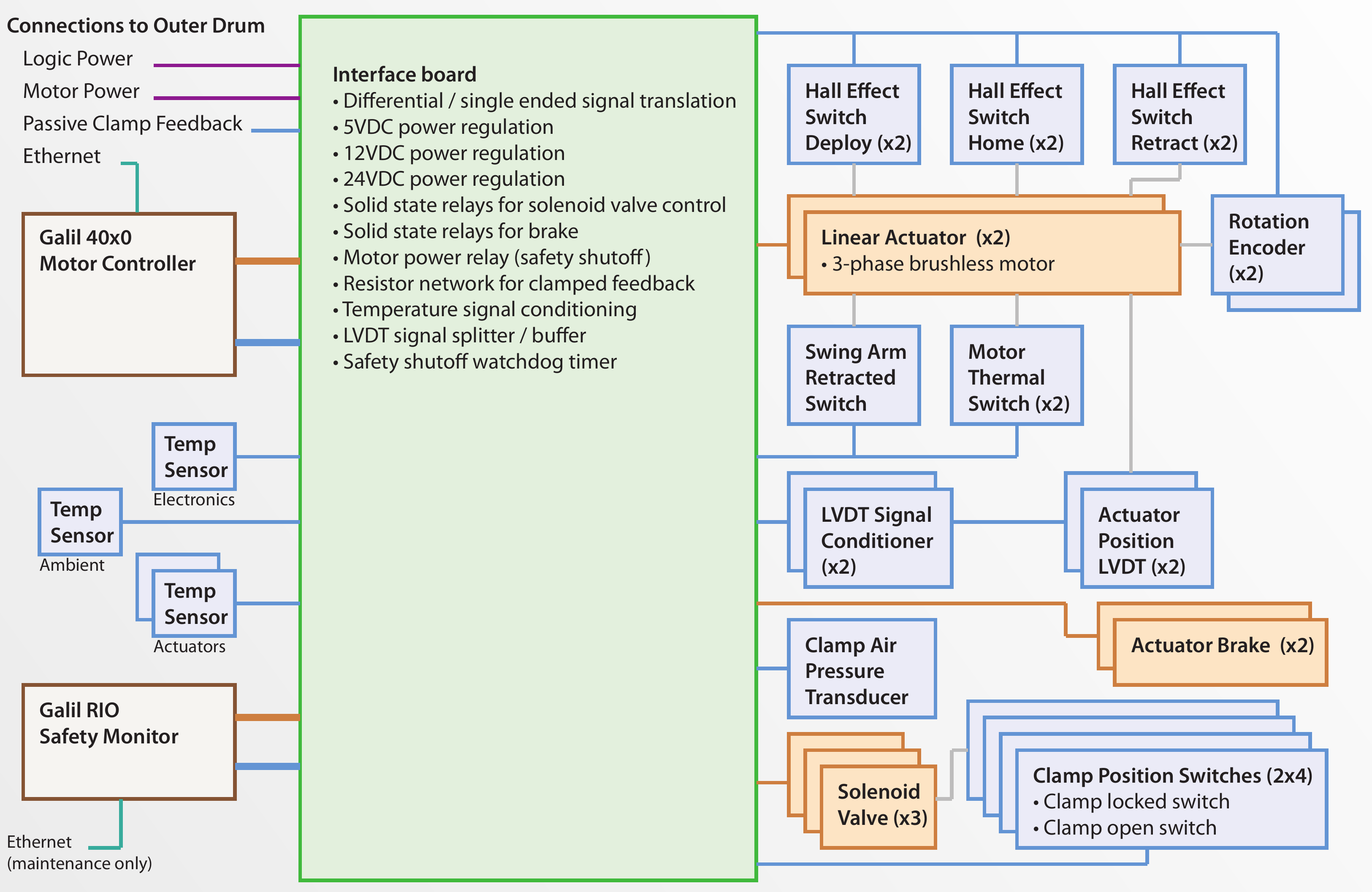}
 \caption{\footnotesize 
 Block diagram of the deployment electronics.
}\label{fig:elec_deploy}
\vskip -0.1in
\end{center}
\end{figure}

The deployment stage electronics are located on the rotating drum portion of K1DM3. These electronics are responsible for mirror deployment and mirror retraction. Power and communication to the deployment stage will be provided through custom
contacts, affixed at two rotational positions on the outer drum. All of the deployment stage electronics will be powered off except when the mirror is being deployed or retracted. An overview of the deployment stage electronics is 
shown in Figure~\ref{fig:elec_deploy}.

Deployment of the mirror is handled by two Exlar linear actuators.  The linear actuators are powered by brushless DC motors.  Each actuator has the following feedback:  a LVDT for absolute position feedback, motor rotation encoder, motor hall effect sensors, temperature sensor, a home location indicator, and switches at the stowed and deployed positions.   Each actuator has a brake.

Kinematic clamping will be done with four pneumatic over-center clamps.  Solenoid valves will control pneumatic cylinders for the clamps.  Feedback switches will verify when the clamps are locked or unlocked.  A pressure transducer will monitor the air supply line for adequate pressure.

Control of the motors, brakes, solenoid valves, and all feedback will be through a Galil 4040 series controller.  Commands to the Galil will be sent over Ethernet.  The Ethernet connection to the Galil will be provided through custom contacts on the drum.

\begin{figure}[h!]
\begin{center}
\includegraphics[width=3.0in,angle=-90]{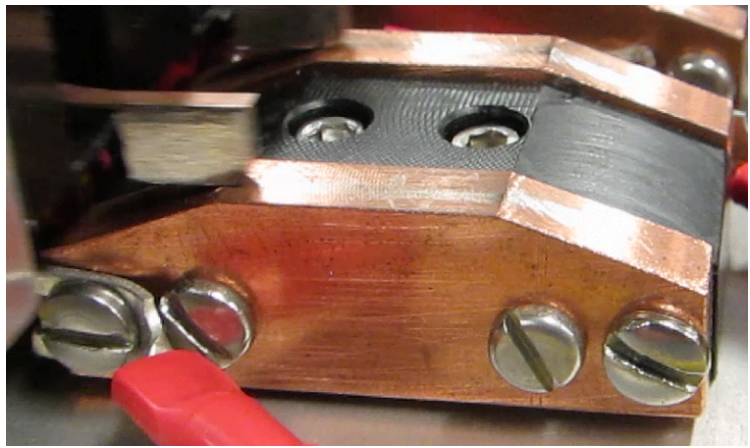}
 \caption[Custom contacts prototype]{\footnotesize 
 Image of a prototype of the custom contacts, brush on ramp.  
 The final system will consist of a series of these contacts coated
 in Ag with a slightly modified brush design.
}\label{fig:contacts}
\vskip -0.1in
\end{center}
\end{figure}

Separate power for the motors and control logic will be provided through the custom contacts. After careful consideration of a slip ring for communications, the team chose to develop a set of custom communications to minimize cost, minimize complexity and to enable easier servicing. 
Our design uses a set of spring contacts (or brushes)
that are pulled across a set of contact ramps 
(Figure~\ref{fig:contacts}).
The contacts are carbon impregnated with silver.   
The ramps will be made from copper and plated with silver. 
We will also likely coat the ramps and brushes with a product
formerly called ProGold.  Our tests show the contact
resistance dropped from 100\,mOhms to 3\,mOhms when using
this product and it is reported to inhibit oxidation.

Power supplies will be located beneath the primary mirror where the heat can be extracted.  Two passive signals for stowed and deployed positions will be passed through the custom contacts. These signals will allow mirror position to be verified without powering up the deployment electronics. 

An independent safety monitoring system will insure the safe operation of deploy and retract movements.  The safety monitor system will be based on a Galil RIO PLC.   The safety monitor system will monitor the position and speed of the deployment stage.  If excessive speed is detected motor power will be cut and the brakes set.   Maximum allowable speed is reduced when approaching stops at either the deployed or retracted position.
The safety monitor also receives a local lockout input to defeat all motion during servicing, and receives the observatory E-stop input which stops all K1DM3 module motion (including the rotator) when the E-stop input is activated.

\subsubsection{Rotation Stage}

\begin{figure}[h!]
\begin{center}
\includegraphics[width=3in]{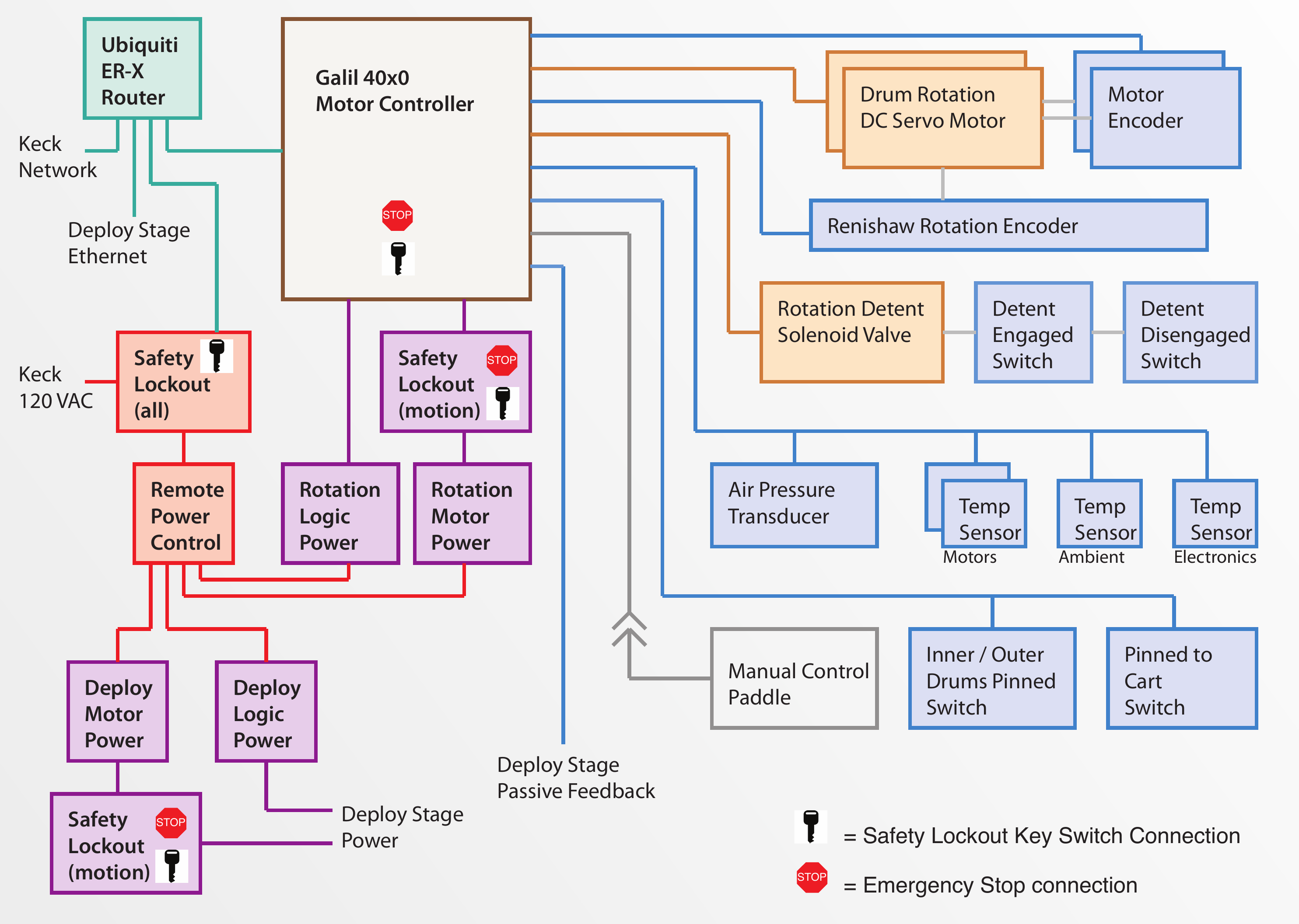}
 \caption{\footnotesize 
 Block diagram of the rotation electronics.
}\label{fig:elec_rotation}
\vskip -0.1in
\end{center}
\end{figure}

The rotation stage electronics controls the angular position of the drum and the detent actuator. An overview of the rotation stage electronics is shown 
in Figure~\ref{fig:elec_rotation}.

Two DC motors will be used to drive the rotation. Rotary 
encoders will provide motor feedback. An absolute position encoder (Renishaw)
and a home switch will provide drum position feedback. Motor temperatures will be monitored 
for system health and safety. 
The motors will be powered down except during movement.

The air actuated kinematic detent mechanism will be controlled via a solenoid valve. Feedback switches will be provided to verify that the detent mechanism is fully engaged or fully retracted.

A Galil controller will be used to control and monitor the motor(s) and detent mechanism. Drum position, motor encoder, detent monitors, and motor temperatures will be fed into the Galil for monitoring and feedback. In addition the Galil will also monitor the mirror retracted/deployed signals from the deployment stage. Communications to the Galil will be via Ethernet. The rotation Galil will remain powered up and will continuously monitor drum position and other feedback sensors.
The rotation stage Galil controller receives a local lockout input to defeat all motion during servicing, and receives the observatory E-stop input which stops all K1DM3 module motion (including the rotator) when the E-stop input is activated.

The Galil controller and power supplies will be located beneath the primary mirror where the heat can be extracted.

\subsubsection{Safety}
\label{sec:elec_safety}

Two levels of hardware safety lockouts will be provided. One level will disable all actuators and motors while leaving all the feedback sensors available. The second level will remove all power from the system.  These satisfy the observatory requirements.

\subsection{Software}

The K1DM3 software, like all WMKO instrument software is based on three
software layers, a low-level server layer, the Keck Task Library (KTL)
layer, and a user interface layer, which provides the graphical user
interfaces (GUIs). 

The low level server layer implements communications and control over the
instrument hardware, and provides a keyword server interface via the KTL
layer. For K1DM3 the hardware motion controllers are two Galil DMC-4040's
and a standalone Galil RIO.  Each of the DMC-4040's is controlled by an
instance of UCO's standard {\it galildisp} daemon, and they are both
part of the single {\it k1dm3} KTL service.

The KTL layer is a standard WMKO software component that is used in
every instrument at the Observatory, allowing client applications
to communicate with any server daemon in a uniform, standardized way.
All data in a server is represented in keyword/value pairs.  Any client
application accesses the KTL layer via KTL's standard library routines.
The "upper half" of the KTL layer is a uniform application programming
interface (API) used in the same way by any application, whereas the
"lower half" of the KTL layer uses one of several different messaging
methods for communicating with KTL servers.  K1DM3 uses KTL/MUSIC,
which uses MUSIC messaging as the transport layer.

The {\it k1dm3} service will support the existing TCS/DCS tertiary
control keywords, and the new keywords defined by the TCSU.
It also has an extensive set of keywords to provide detailed
engineering support, and detailed feedback about the status
of all controlled and monitored elements.




\acknowledgments     
 
This material is based upon work supported in part by the National Science Foundation under Grant No. AST-1337609. We also acknowledge cost-sharing by the W.M. Keck Observatory, UC Observatories, and UC Santa Cruz. Undergraduate Alex Tripsas is also supported by an REU supplemental to the NSF grant.



\begin{thebibliography}{10}

\bibitem{k1dm3_requirements}
{WMKO}, ``Requirements for {K1DM3}: The {Keck} {I} deployable tertiary
  mirror,'' Tech. Rep. v3.3, WMKO (March 2016).

\bibitem{k1dm3_mirror_spec}
{K1DM3 Team}, ``{K1DM3 Design Note}: Mirror specifications for the {K1DM3}
  project,'' Tech. Rep. v1.3, UCO (September 2014).

\bibitem{k1dm3_asbuilt}
{K1DM3 Team}, ``{K1DM3 Design Note}: Elevation axis of the {Keck} 1
  telescope,'' Tech. Rep. v1.5, UCO (September 2015).

\bibitem{k1dm3_positioning}
{K1DM3 Team}, ``{K1DM3 Design Note}: Positioning of {M3} for the {K1DM3}
  project,'' Tech. Rep. v2.4, UCO (August 2014).

\bibitem{k1dm3_zygo}
{K1DM3 Team}, ``{K1DM3 Design Note}: Zygo quote for the {K1DM3} mirror,'' Tech.
  Rep. v1.3.1, UCO (September 2015).

\bibitem{k1dm3_ADC}
{K1DM3 Team}, ``{K1DM3 Design Note}: Vignetting with {K1DM3} at the cassegrain
  focus, with {ADC},'' Tech. Rep. v1.1, UCO (September 2015).

\bibitem{kotn_579}
{McBride}, D. and {Hudek}, S., ``{KOTN} 579; etching zerodur with hydrofluoric
  acid vs. ammonium bifluoride,'' tech. rep., WMKO (July 2011).

\bibitem{kotn_580}
{McBride}, D. and {Hudek}, S., ``{KOTN} 580: Mirror segment repair project,
  adhesive selection for axial insert repairs,'' tech. rep., WMKO (September
  2011).

\bibitem{kotn_803}
{McBride}, D., ``{KOTN} 803: Axial insert load vs displacement,'' tech. rep.,
  WMKO (April 2015).

\bibitem{kotn_804}
{McBride}, D., ``{KOTN} 804: Axial insert adhesive pad size vs strength,''
  tech. rep., WMKO (April 2015).

\bibitem{kotn_819}
{McBride}, D., ``{KOTN} 819: E-120hp adhesive hardness vs cure time,'' tech.
  rep., WMKO (July 2015).

\bibitem{kotn_823}
{McBride}, D., ``{KOTN} 823: Axial insert strength versus adhesive cure time,''
  tech. rep., WMKO (October 2015).

\bibitem{kotn_824}
{McBride}, D., ``{KOTN} 824: Axial insert and radial pad qualification tests,''
  tech. rep., WMKO (October 2015).

\bibitem{slocum92}
{Slocum}, A.,  [{\em {A Precision Machine
  Design}}{\nolinebreak\hspace{0.1em}]}, Society of Manufacturing Engineers,
  Dearborn, MI (1992).

\bibitem{k1dm3_ODF}
{K1DM3 Team}, ``{K1DM3 Design Note}: Outer drum fitting alignment,'' Tech. Rep.
  v1.3, UCO (February 2016).

\end{thebibliography}

\end{document}